\documentclass[structabstract]{aa}
\usepackage{graphicx}
\usepackage{txfonts}
\usepackage{verbatim}
\usepackage{natbib}
\bibpunct{(}{)}{;}{a}{}{,} 

\begin{document}

\title{Doppler images of II Pegasi for 2004--2010\thanks{Based on observations made with the Nordic Optical Telescope, operated
on the island of La Palma jointly by Denmark, Finland, Iceland,
Norway, and Sweden, in the Spanish Observatorio del Roque de los
Muchachos of the Instituto de Astrofisica de Canarias. }}
\author{T. Hackman \inst{1,} \inst{2} 
\and M.J. Mantere \inst{1}
\and M. Lindborg \inst{3,} \inst{1}
\and I. Ilyin \inst{4}
\and O. Kochukhov \inst{5}
\and N. Piskunov \inst{5}
\and I. Tuominen \inst{1,}\inst{6}}

\institute{Department of Physics, P.O. Box 64 , 
FI-00014 University of Helsinki, Finland \\
\email{Thomas.Hackman@helsinki.fi}
\and Finnish Centre for Astronomy with ESO, University of Turku, 
V\"{a}is\"{a}l\"{a}ntie 20, FI-21500 Piikki\"{o}, Finland
\and Nordic Optical Telescope, 38700 Santa Cruz de la Palma, Spain
\and Leibniz-Institut f\"{u}r Astrophysik Potsdam, An der Sternwarte 16, 14882 Potsdam,
 Germany
\and Department of Physics and Astronomy, Uppsala University, Box 516, 
SE-751 20 Uppsala, Sweden
\and Deceased}

\date{Received / accepted}

\abstract{} {We study the spot activity of \object{II Peg} during the
  years 2004--2010 to determine long- and short-term changes
  in the magnetic activity. In a previous study, we detected a
  persistent active longitude, as well as major changes in the spot
  configuration occurring on a timescale of 
  shorter than a year. The
  main objective of this study is to determine whether the same
  phenomena persist in the star during 
these six years of spectroscopic monitoring.}
{The observations were collected with the high-resolution SOFIN
  spectrograph at the Nordic Optical Telescope. The temperature maps
  were calculated using a Doppler imaging code based on Tikhonov
  regularization.}  
{We present 12 new temperature maps that show spots distributed mainly
  over high and intermediate latitudes. In each 
  image, 1-3 main active
  regions can be identified. The activity level of the star is clearly
  lower than during our previous study 
  for the years 1994--2002. In contrast to the
  previous observations, we detect no clear drift of the active
  regions with respect to the rotation of the star. 
}
{Having shown a systematic longitudinal drift of the spot-generating
  mechanism during 1994--2002, the star has clearly switched to a
  low-activity state for 2004--2010, during which the spot locations
  appear more random over phase space. It could be that the star is
near to a
minimum of its activity cycle. 
}


\maketitle

\section{Introduction}

\object{II Peg} (\object{HD 224085}) is an extremely active RS CVn star, 
exhibiting both strong chromospheric and coronal activity, 
in addition to photometric variability 
caused by spots. It has had one of the highest observed photometric 
amplitudes \citep{doyle1989} and is also claimed to be the brightest X-ray 
source within 50 pc \citep{makarov2003}. 
More recent studies of II Peg include estimates of its differential rotation
using ground-based 
\citep{roettenbacher2011} and satellite observations 
\citep{siwak2010}. Both studies indicate a weak solar-type differential 
rotation. The latter study 
based on data from the MOST satellite
confirmed previous findings that flares must be related to active regions,
since they are
more frequently observed when the most spotted hemisphere is visible
\citep[e.g.][]{mohin1993,teriaca1999,frasca2008}.

\object{II Peg} has been spectroscopically monitored for nearly
20 years with the SOFIN spectrograph at the Nordic Optical
Telescope (La Palma, Spain). Its observations have 
enabled both Doppler imaging and long-term studies
of the spot activity to be performed
\citep{berdyugina1998,berdyugina1999,
  lindborg2011}. A summary of these results can be found in our
earlier paper \citet{lindborg2011}, which was based on
observations from 1994 to 2002.
In these Doppler images, 
there were usually two active longitudes, one
of which was persistent and stronger almost throughout the 
nine years of spectroscopic monitoring.
Major changes, occurring on a timescale of shorter than one
year, could also be
seen. For instance, the spot activity could switch to another,
weaker, longitude, for a short period of
time. These shifts are similar to the 'flip-flops' described by Berdyugina
and collaborators \citep{berdyugina1998,BT98,berdyugina1999}, except that they
were not found to follow any periodicity.
In addition to the short-term shifts of the active 
longitudes, we also discovered a drift in
the active regions with respect to the
orbital rotation frame, indicating that the spot-generating structure
rotated slightly faster than the
tidally locked binary system.

Rapidly rotating late-type stars with deep convective envelopes
are expected to exhibit very little
differential rotation, because it is theoretically predicted to be suppressed
in the rapid rotation regime \citep[e.g.][]{KR99}; for \object{II Peg},
this has also been observationally confirmed 
\citep{siwak2010,roettenbacher2011}.
However, several stars have been 
found to contradict
this prediction \citep[e.g.][]{hackman2001,jeffers2008,frasca2011}. 
Furthermore it is unclear whether 
surface differential rotation can be recovered by following the motion of
large spots \citep{korhonen2011}.
Nevertheless, in the case of suppressed differential rotation, the
dynamos working in rapidly rotating stars
are expected to be of the $\alpha^2$-type,
the magnetic field generation only
being due to the inductive action of
convective turbulence.
According to both linear
\citep[e.g.][]{krause1980meanfield} and nonlinear solutions
\citep[e.g.][]{MBBT95} of the $\alpha^2$-dynamo equations, the
nonaxisymmetric modes become more easily excited in the rapid rotation
regime. The 
$m\!=\!\!1$ mode, representing an azimuthally varying field
changing sign once over the full longitude span, is commonly the
preferred field configuration. This is unsurprising, as the largest
scale mode is always the least affected by diffusive effects.

The nonaxisymmetric modes turn out to be waves migrating in the azimuthal
direction, not necessarily having the rotation period of the star
\citep[e.g.][]{krause1980meanfield}. Both slower and faster dynamo
waves can occur, depending for example on the profile and properties
of the turbulent transport coefficients. The faster waves with dipole
symmetry (S1) were found to be preferred  in linear models with
simple profiles \citep{krause1980meanfield}, and slower waves with
quadrupolar symmetry (A1) in more complicated nonlinear models 
when solving
for the dynamics \citep{tuominen2002starspot}. From the viewpoint
of dynamo theory, two migratory active longitudes are thus
an expected result. We, therefore, interpret the behaviour seen in
\object{II Peg} during 1994--2002 as
a manifestation of such a
dynamo wave. The objective of the present paper was to inspect whether
the azimuthal dynamo wave persisted on the star in the most recent
observational data of the object.

\begin{table}
\caption{Summary of observations with NOT.$^{\mathrm{a}}$}
\centering
\begin{tabular}{ccccccc}
\hline \hline
Season  & $t_{\min}$      & $t_{\max}$       & $S/N$ & $n_\phi$ & $f_\phi$ & $d$\\ 
        &               &                &       &         &         &$\times 10^{-3}$\\
\hline
Aug04   &  53216.6    & 53228.6     & 244     &  8 &  75\% & 5.70\\
Dec04   &  53370.4    & 53372.4     & 177     &  3 &  30\% & 6.19\\
Jul05   &  53567.7    & 53575.6     & 218     &  7 &  56\% & 5.94\\
Nov05   &  53685.5    & 53695.5     & 197     &  5 &  38\% & 6.73\\
Sep06   &  53978.5    & 53992.7     & 259     & 11 &  72\% & 6.10\\
Dec06   &  54071.5    & 54078.4     & 166     &  7 &  63\% & 6.66\\
Jul07   &  54300.7    & 54309.7     & 264     & 10 &  81\% & 5.84\\
Nov07   &  54427.4    & 54437.4     & 238     &  7 &  63\% & 6.19\\
Sep08   &  54717.6    & 54723.5     & 304     &  4 &  40\% & 5.37\\
Dec08   &  54809.4    & 54815.5     & 263     &  6 &  59\% & 7.19\\
Aug09   &  55069.7    & 55081.7     & 291     & 12 &  87\% & 6.82\\
Dec09   &  55193.4    & 55201.4     & 283     &  8 &  68\% & 6.02\\
\hline
\end{tabular}
\begin{list}{}{}
\item[$^{\mathrm{a}}$] Season, $HJD-2400000$ 
of first ($t_{\min}$) and last ($t_{\max}$) 
observation, mean signal-to-noise ratio ($S/N$), number of observed 
phases ($n_\phi$), estimated phase coverage ($f_\phi$), and deviation ($d$) of 
the Doppler imaging solution.
\end{list}
\label{obssum}
\end{table}

\begin{table*}
\caption{Observations with NOT.$^{\mathrm{b}}$} 
\centering
\begin{tabular}{cccr|cccr|cccr}

\hline \hline
Date & HJD & $\phi$ & $S/N$ & Date & HJD & $\phi$ &  $S/N$
 & Date & HJD & $\phi$ &  $S/N$
  \\
(dd/mm/yyyy) & & &     &  (dd/mm/yyyy)  &  & &   
& (dd/mm/yyyy) & & & \\
\hline
30/07/2004  & 3216.618 & 0.379 & 190 & 11/09/2006  & 3989.704 & 0.348 & 314 &
11/09/2008  & 4720.555 & 0.035 & 290 \\
31/07/2004  & 3217.646 & 0.532 & 291 & 12/09/2006  & 3990.649 & 0.488 & 404 &
13/09/2008  & 4722.626 & 0.343 & 320 \\
01/08/2004  & 3218.648 & 0.681 & 249 & 13/09/2006  & 3991.680 & 0.642 & 169 & 
 14/09/2008  & 4723.544 & 0.480 & 306 \\
02/08/2004  & 3219.644 & 0.829 & 247 & 14/09/2006  & 3992.680 & 0.790 & 213 &
 08/12/2008  & 4809.393 & 0.247 & 383 \\
03/08/2004  & 3220.653 & 0.979 & 201 & 01/12/2006  & 4071.466 & 0.507 & 198 &
09/12/2008  & 4810.421 & 0.399 &  104 \\
05/08/2004  & 3222.612 & 0.271 & 227 & 03/12/2006  & 4072.526 & 0.665 & 149 &
 10/12/2008  & 4811.478 & 0.557 & 222 \\
10/08/2004  & 3227.721 & 0.031 & 270 & 04/12/2006  & 4074.413 & 0.945 & 183 &
 11/12/2008  & 4812.439 & 0.700 & 260 \\
11/08/2004  & 3228.615 & 0.164 & 278 & 05/12/2006  & 4075.472 & 0.103 & 159 &
 13/12/2008  & 4814.437 & 0.997 & 340 \\
30/12/2004  & 3370.395 & 0.248 & 237 & 06/12/2006  & 4076.442 & 0.247 & 188 &
 14/12/2008  & 4815.474 & 0.151 & 268 \\
31/12/2004  & 3371.377 & 0.394 & 192 & 07/12/2006  & 4077.427 & 0.394 & 151 &
 26/08/2009  & 5069.676 & 0.954 & 302 \\
01/01/2005  & 3372.382 & 0.544 &  103 & 08/12/2006  & 4078.415 & 0.540 &  136 &
 27/08/2009  & 5070.687 & 0.105 & 290 \\
16/07/2005  & 3567.704 & 0.591 & 210 & 19/07/2007  & 4300.696 & 0.597 & 233 &
28/08/2009  & 5071.725 & 0.259 & 184 \\
17/07/2005  & 3568.713 & 0.741 & 205 & 20/07/2007  & 4301.710 & 0.747 & 266 & 
 29/08/2009  & 5072.697 & 0.404 & 320 \\
18/07/2005  & 3569.728 & 0.892 & 207 & 21/07/2007  & 4302.674 & 0.891 & 276 & 
30/08/2009  & 5073.708 & 0.554 & 186 \\
19/07/2005  & 3570.644 & 0.028 & 228 & 22/07/2007  & 4303.630 & 0.033 & 268 &
 01/09/2009  & 5075.557 & 0.829 & 254 \\
20/07/2005  & 3571.661 & 0.179 & 234 & 23/07/2007  & 4304.673 & 0.188 & 242 &
 02/09/2009  & 5076.682 & 0.996 & 364 \\
23/07/2005  & 3574.675 & 0.627 & 217 & 24/07/2007  & 4305.705 & 0.341 & 299 &
 03/09/2009  & 5077.634 & 0.138 & 411 \\
24/07/2005  & 3575.649 & 0.772 & 230 & 25/07/2007  & 4306.643 & 0.481 & 314 &
 04/09/2009  & 5078.743 & 0.303 & 333 \\
11/11/2005  & 3685.505 & 0.109 &  122 & 26/07/2007  & 4307.705 & 0.639 & 244 &
 05/09/2009  & 5079.676 & 0.442 & 346 \\
11/11/2005  & 3686.479 & 0.254 & 146 & 27/07/2007  & 4308.649 & 0.779 & 275 &
 06/09/2009  & 5080.672 & 0.590 & 348 \\
17/11/2005  & 3692.484 & 0.147 & 265 & 28/07/2007  & 4309.698 & 0.935 & 222 &
 07/09/2009  & 5081.740 & 0.748 & 164 \\
19/11/2005  & 3693.548 & 0.305 & 226 & 22/11/2007  & 4427.437 & 0.445 & 241 &
 27/12/2009  & 5193.360 & 0.348 & 271 \\
20/11/2005  & 3695.471 & 0.591 & 224 & 25/11/2007  & 4430.472 & 0.896 & 152 &
 28/12/2009  & 5194.381 & 0.500 & 264 \\
30/08/2006  & 3978.502 & 0.682 & 214 & 26/11/2007  & 4431.475 & 0.045 & 236 &
 29/12/2009  & 5195.371 & 0.647 & 366 \\
01/09/2006  & 3979.654 & 0.853 & 244 & 27/11/2007  & 4432.454 & 0.191 & 343 &
 31/12/2009  & 5197.399 & 0.949 & 366 \\
02/09/2006  & 3980.686 & 0.007 & 240 & 30/11/2007  & 4435.317 & 0.617 &  115 &
 01/01/2010  & 5198.411 & 0.099 & 310 \\
05/09/2006  & 3983.672 & 0.451 & 156 & 01/12/2007  & 4436.470 & 0.788 & 294 &
 02/01/2010  & 5199.376 & 0.243 & 259 \\
06/09/2006  & 3984.731 & 0.608 & 212 & 02/12/2007  & 4437.448 & 0.934 & 290 &
 03/01/2010  & 5200.363 & 0.389 & 213 \\
07/09/2006  & 3985.676 & 0.749 & 340 & 08/09/2008  & 4717.606 & 0.597 & 303 &
 04/01/2010  & 5201.385 & 0.541 & 217 \\
09/09/2006  & 3987.705 & 0.051 & 338 & & & & & & & & \\
\hline
\end{tabular}
\begin{list}{}{}
\item[$^{\mathrm{b}}$] The heliocentric Julian date is given as
HJD$-2450 000$ and the $S/N$-ratio is for 
a wavelength region 
centred on
5630 \AA.
\end{list}
\label{obsful}
\end{table*}

\section{Observations}
\label{observations}

Our observations were made using the SOFIN high-resolution
\'echelle spectrograph at the 2.56m Nordic Optical Telescope (NOT),
La Palma, Spain. The data were acquired with the second camera
equipped with a Loral CCD detector with 2048x2048 pixels. This provides a 
spectral resolution of $R \approx 70000$.

A total of 12 sets of high-resolution spectra of II
Peg were measured in 2004--2010.
The signal-to-noise ratio ($S/N$) 
of the observations was usually around 200--300. The number of 
observed
phases ranged from 3 to 12. 

In general, observations of about ten
evenly distributed rotation phases is 
considered optimal for Doppler imaging \citep[e.g.][]{vogt1987}. The spatial 
resolution of the Doppler imaging depends on the phase coverage, but the phase 
coverage is more important for determining the right spot latitude 
than the longitude. In the case of gaps in the observations, the reliability
of the spot positions depends on the visibility of the spots at the times
of observations. This means that
observations with insufficient phase coverage can still be
very useful when studying e.g. active longitudes.
We estimated the fraction $f$ of the rotation phases covered by 
the observations by assuming that each observation
covers $\pm$0.05 of the rotation period. Thus, ten
observations
uniformly distributed in phase 
would give a 100\% coverage. 

The spectral regions 5392.3 -- 5395.1 \AA, 5524.7 -- 5527.3 \AA, and 
5633.2 -- 5634.6 \AA\,were chosen for Doppler imaging. The regions contain
relatively unblended lines of different atoms, ionization 
states, and strengths.
Owing to the diversity of the line parameters, the Doppler imaging 
solution becomes more reliable.

The ephemeris used to calculate the phases 
\begin{equation}  T_\mathrm{conj}=2449582.9268+6.724333E \end{equation}
was taken from \citet{berdyugina1998}. A summary of the observations is 
given in Table~\ref{obssum}. A more complete listing of
the heliocentric Julian dates, phases
calculated according to the ephemeris given above, and the $S/N$
of each observation is given in Table~\ref{obsful}.

The spectral observations were reduced with the 4A software system
\citep{ilyin2000}. Bias, cosmic ray, flat-field and scattered-light
corrections, wavelength calibration and normalization, and corrections for
the motion of the Earth as well as the orbital motion of the binary system
were included in the reduction process.
For the orbital motion, we used the solution presented by \cite{berd98a}. The 
continuum normalization was done in two steps. The spectral orders were first 
normalized by a polynomial continuum fit of third degree. In rapidly
rotating stars, lines are blended, and there may not be any real
continuum within a spectral interval. Therefore, an additional continuum
correction for each wavelength interval used for Doppler imaging was
done by comparing the seasonal average observed profile and a
synthetic line profile. Near-continuum points were used for a first or
second degree polynomial fit to correct the normalized flux level.

\section{Doppler imaging}
\label{DI}

The same Doppler imaging method as that applied
 in \citet{lindborg2011} was used to
calculate temperature maps for II Peg. The inversion technique 
was thoroughly described in several papers
\citep{piskunov1991,hackman2001, lindborg2011}. The main difference
with our previous analysis was that we used the new MARCS model
atmospheres \citep{marcs} for the line profile calculations.

\subsection{Stellar and spectral parameters}

Several different sets of stellar parameters have previously
been suggested for
II Peg.  The choice of parameters is far from trivial in the sense
that different parameters have very similar effects on the spectral
lines and good fits can be obtained using different combinations of
stellar and spectral parameters. For example, changing the value of
the microturbulence will cause shifts in the average effective
temperature $T_\mathrm{eff}$ of the solution. However,
the effect of the spots on the surface is, up to 
first order, the same independent of the chosen
set of parameters:
cool spots cause ``emission bumps'' in the
photospheric spectral lines. 

For the rotational velocity, macroturbulence, 
rotation period, and inclination angle, 
we used the values adopted by \citet{berdyugina1998}.
An unspotted star with $T_\mathrm{eff}=4750$ was 
used as an initial guess of the surface temperature.
Other parameters were chosen by comparing the mean seasonal observations
with synthetic spectra of this unspotted star. 
In practice, the parameters were optimized such
that the spectra, excluding the ``emission bumps'' caused by spots, would 
fit a star with $T_\mathrm{eff} \approx$ 4750 K.

Since our aim was to study the spot activity and not to make
an absolute determination of 
the stellar parameters, it was certainly sufficient to test different sets of
parameters and choose the one providing the best fit to an assumed
unspotted star.  Using the MARCS model atmospheres \citep{marcs}, we
found that the best starting point was provided by
the stellar parameters
suggested by \citet{ottmann1998}. The adopted parameters are given in
Table~\ref{sparam}. 

The parameters for the spectral lines were obtained from VALD
\citep{vald2}. 
To optimize the fit to the observations, the
$\log gf$-value of some lines was changed.  The spectral parameters
for the most important lines are given in Table~\ref{speparam}. The
full line synthesis included 151
lines.

The need to adjust the spectral parameters could indicate 
that the element abundances differ from those of
the standard MARCS model. It should be emphasized that
these kinds of corrections are necessary in Doppler imaging 
to reduce the systematic errors caused by
discrepancies between the synthetic spectra and observations.

\subsection{Inversion procedure}

A table of line profiles was calculated using MARCS plane-parallel
atmosphere models with $T_\mathrm{eff}$ ranging from 3200 K to 6000 K
and the stellar parameters listed in Table~\ref{sparam}. The local
line profiles table and the observations were used as input to
the inversion code.

\begin{figure*}
\begin{center}

\vspace{-2.8cm}
\includegraphics[width=3.5in]{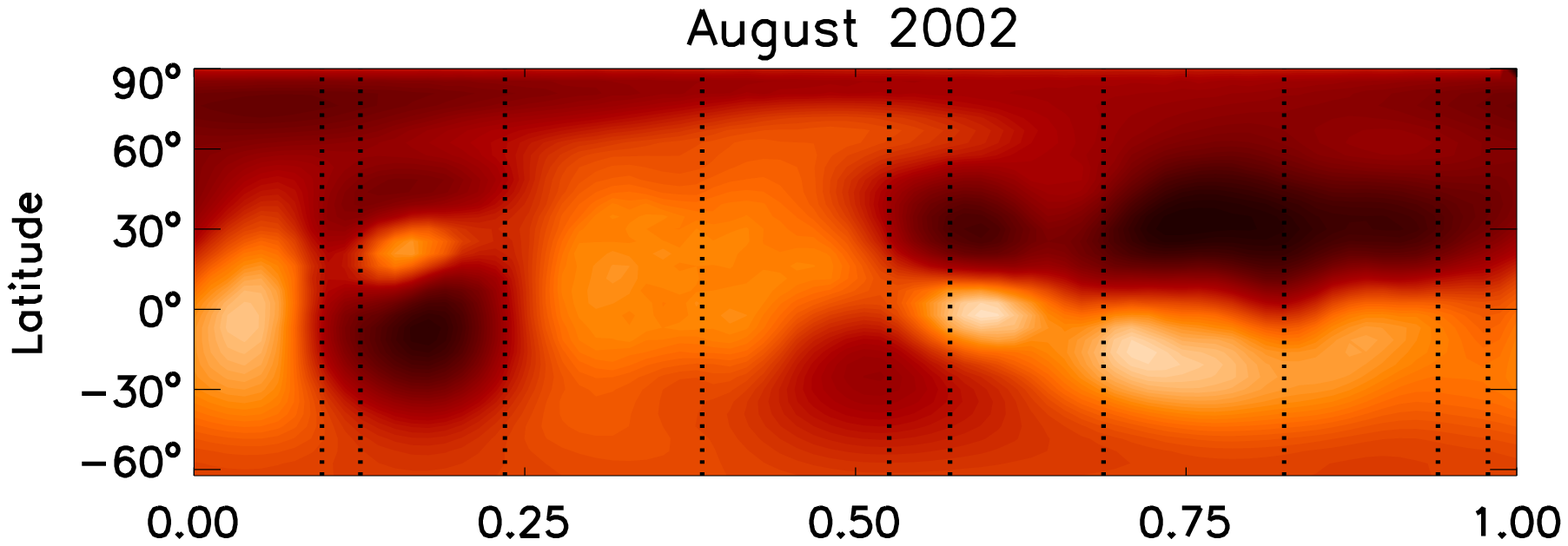}
\includegraphics[width=3.5in]{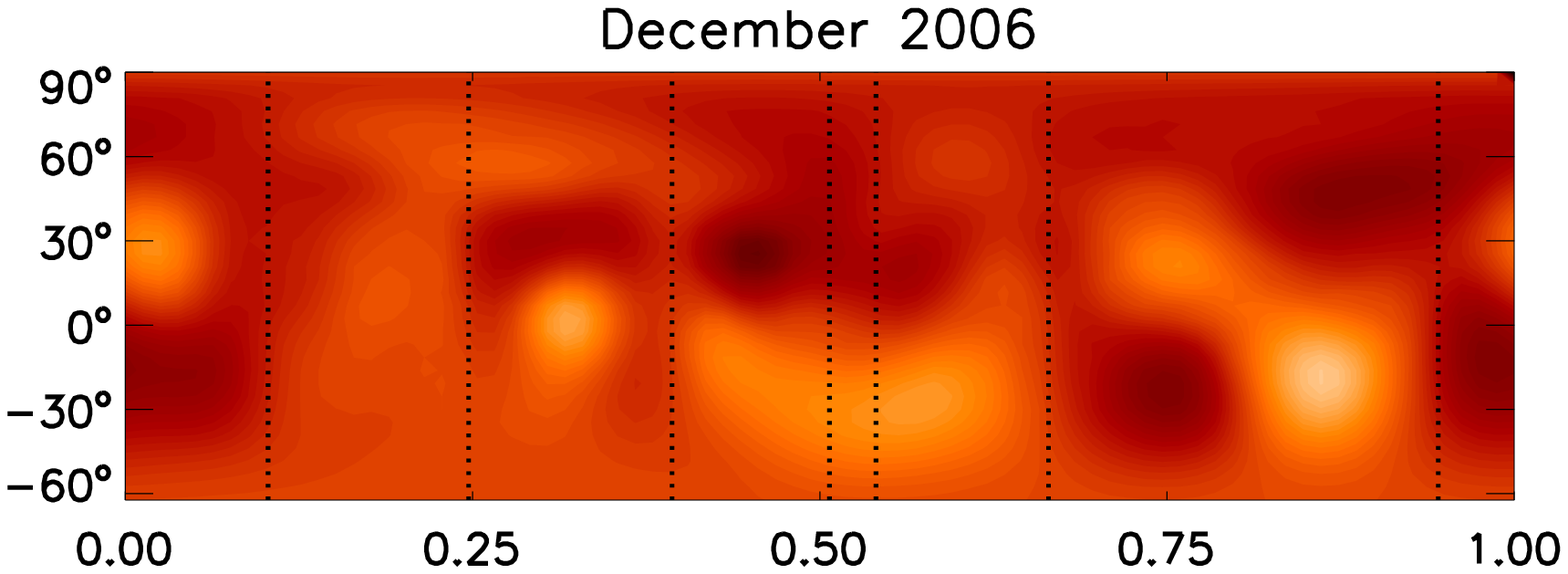}

\vspace{-3.1cm}
\includegraphics[width=3.5in]{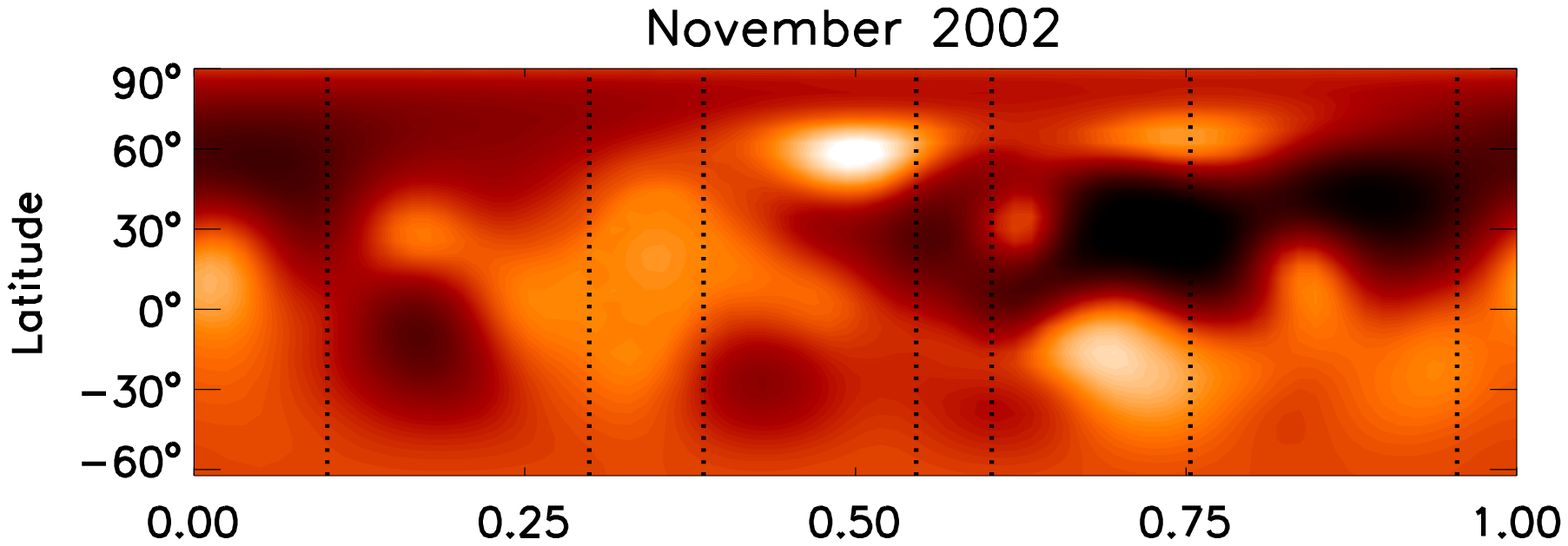}
\includegraphics[width=3.5in]{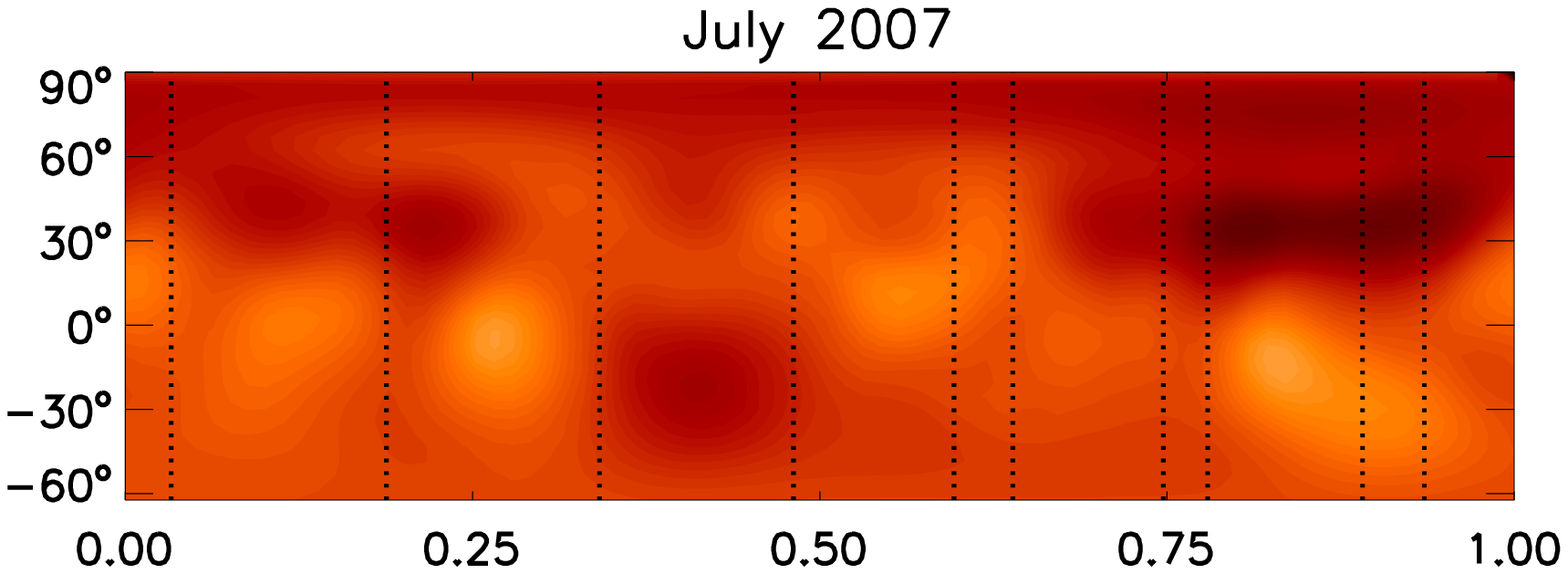}

\vspace{-3.1cm}
\includegraphics[width=3.5in]{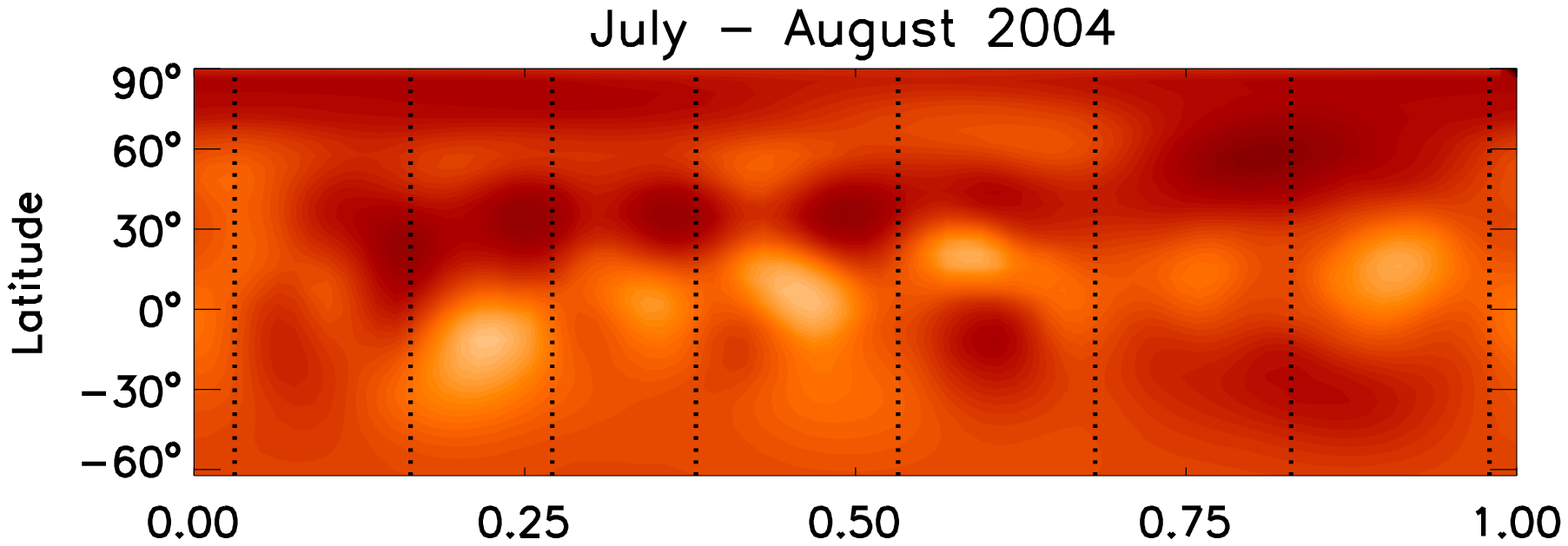}
\includegraphics[width=3.5in]{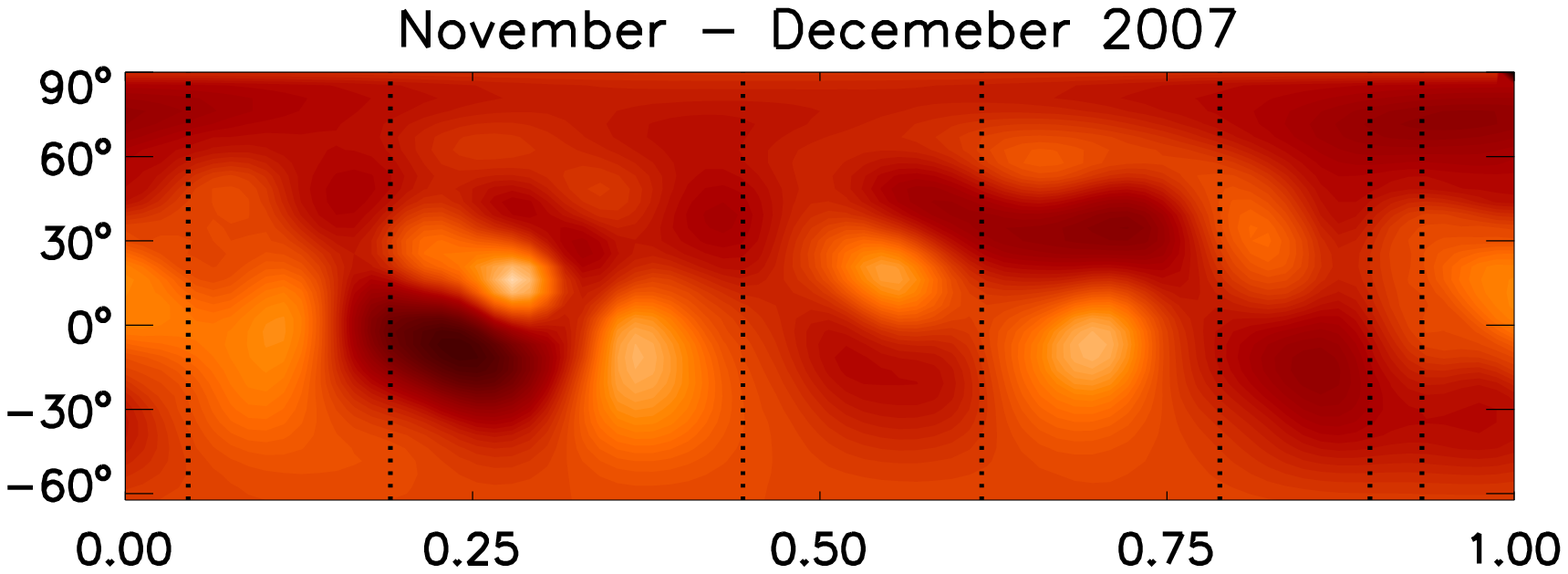}

\vspace{-3.1cm}
\includegraphics[width=3.5in]{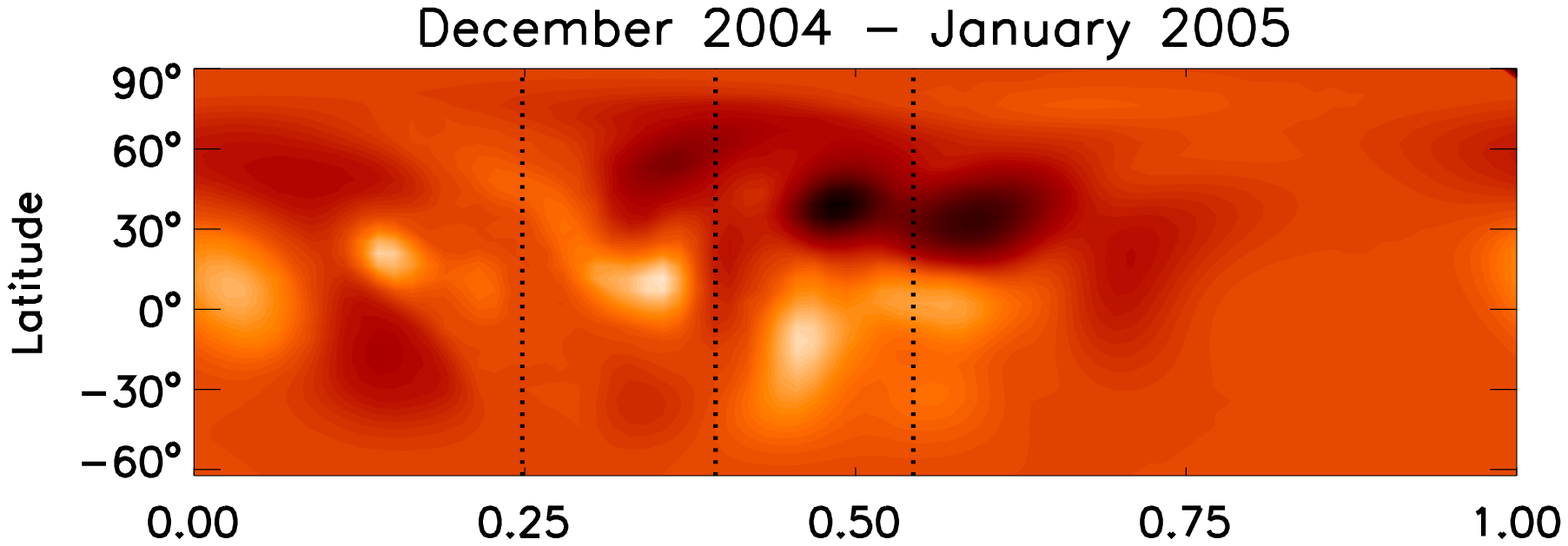}
\includegraphics[width=3.5in]{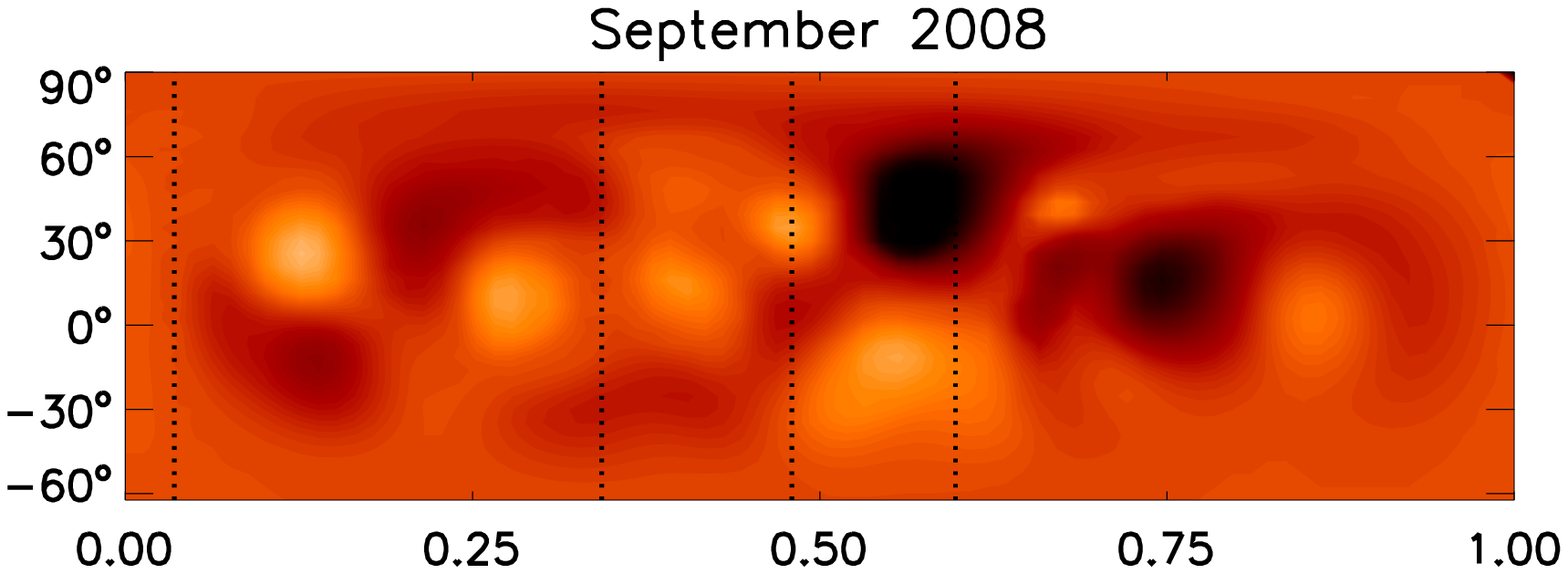}

\vspace{-3.1cm}
\includegraphics[width=3.5in]{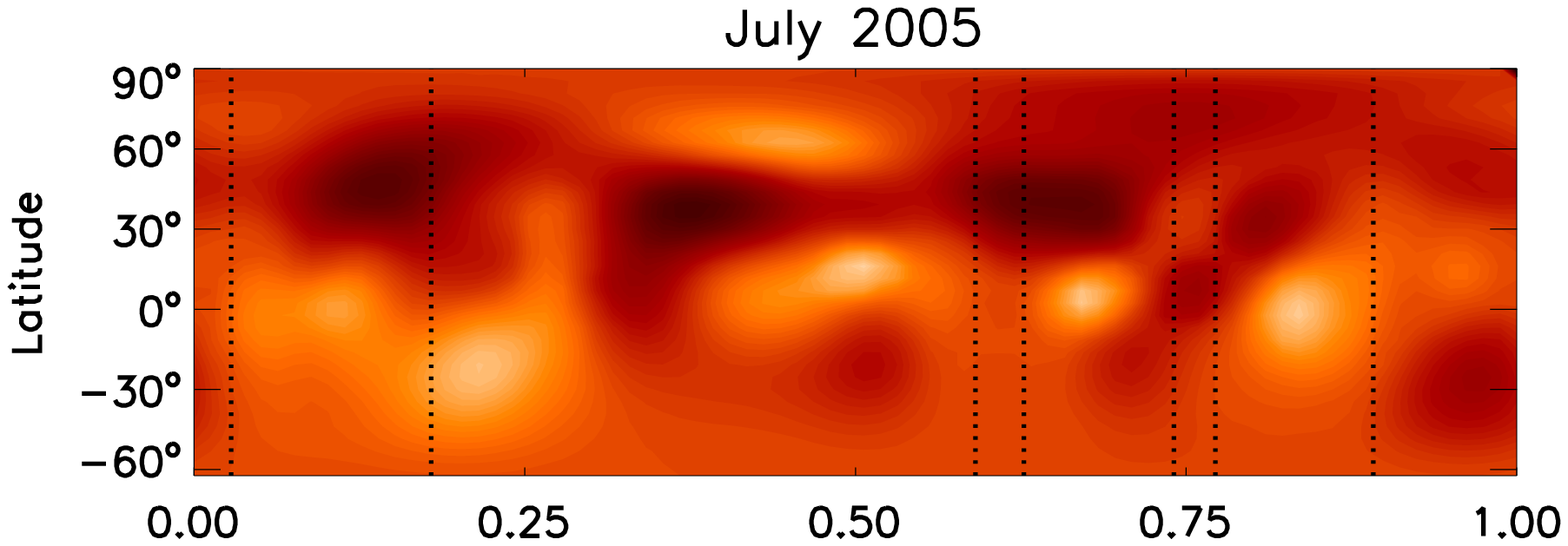}
\includegraphics[width=3.5in]{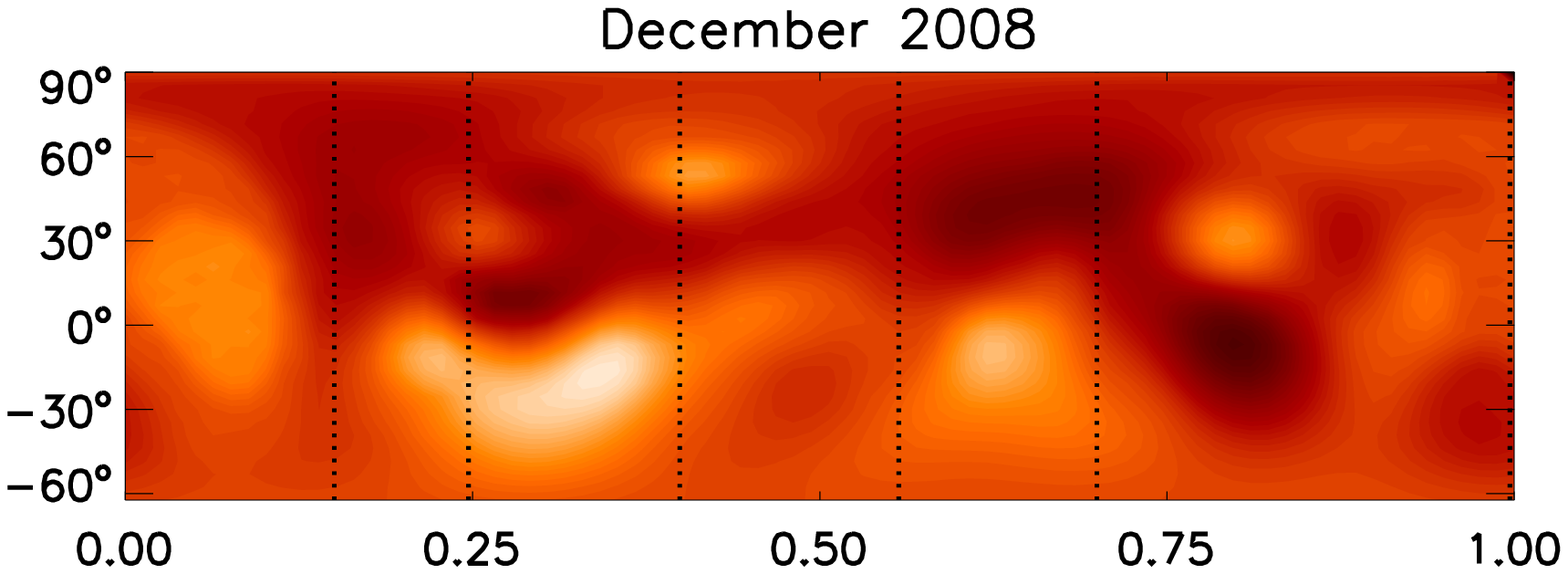}

\vspace{-3.1cm}
\includegraphics[width=3.5in]{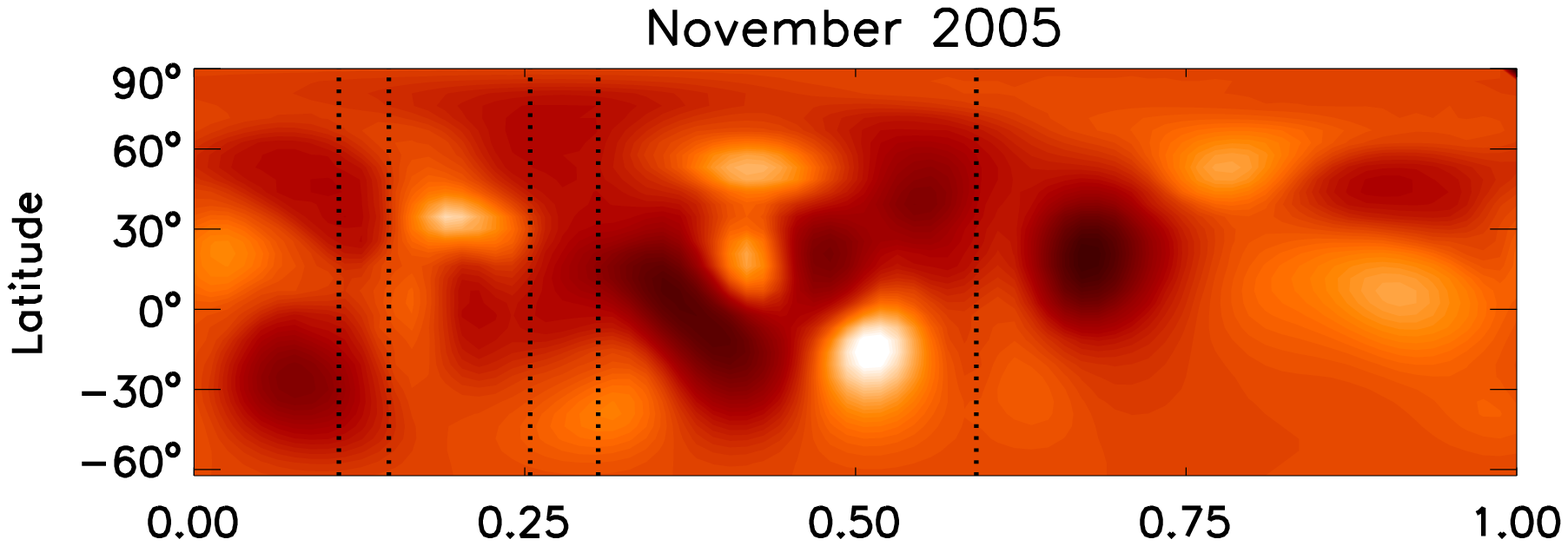}
\includegraphics[width=3.5in]{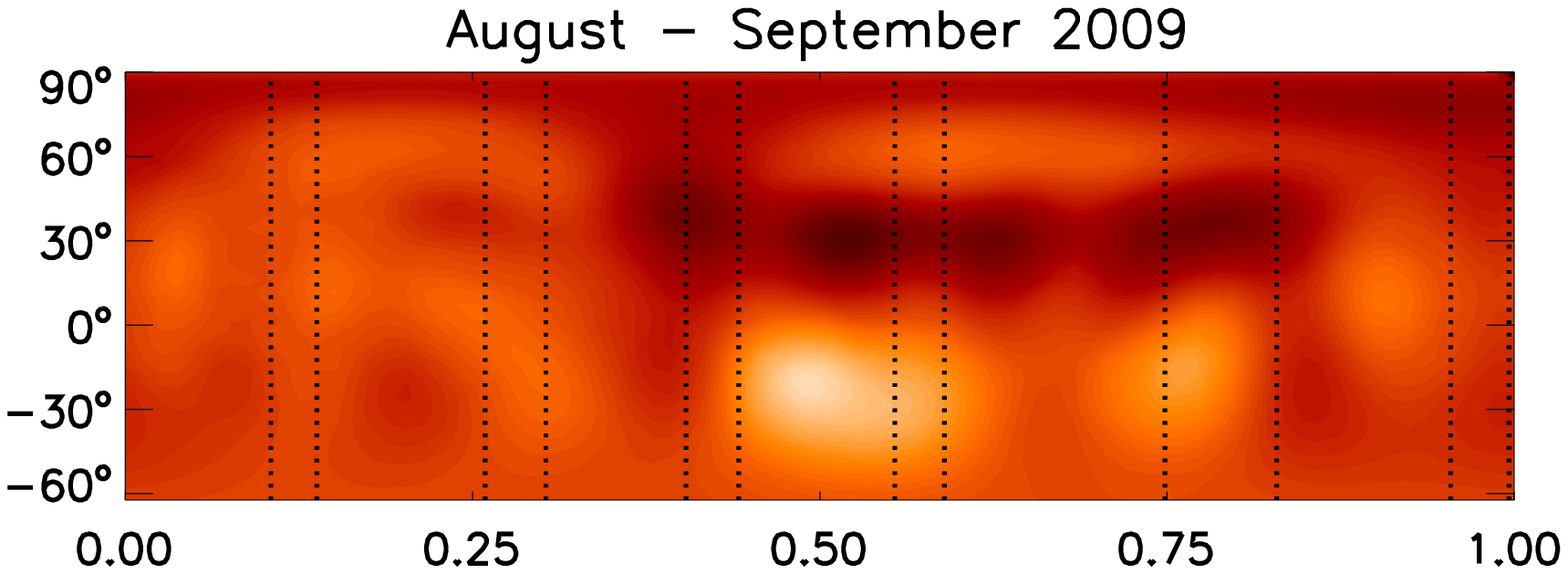}

\vspace{-2.8cm}
\includegraphics[width=3.5in]{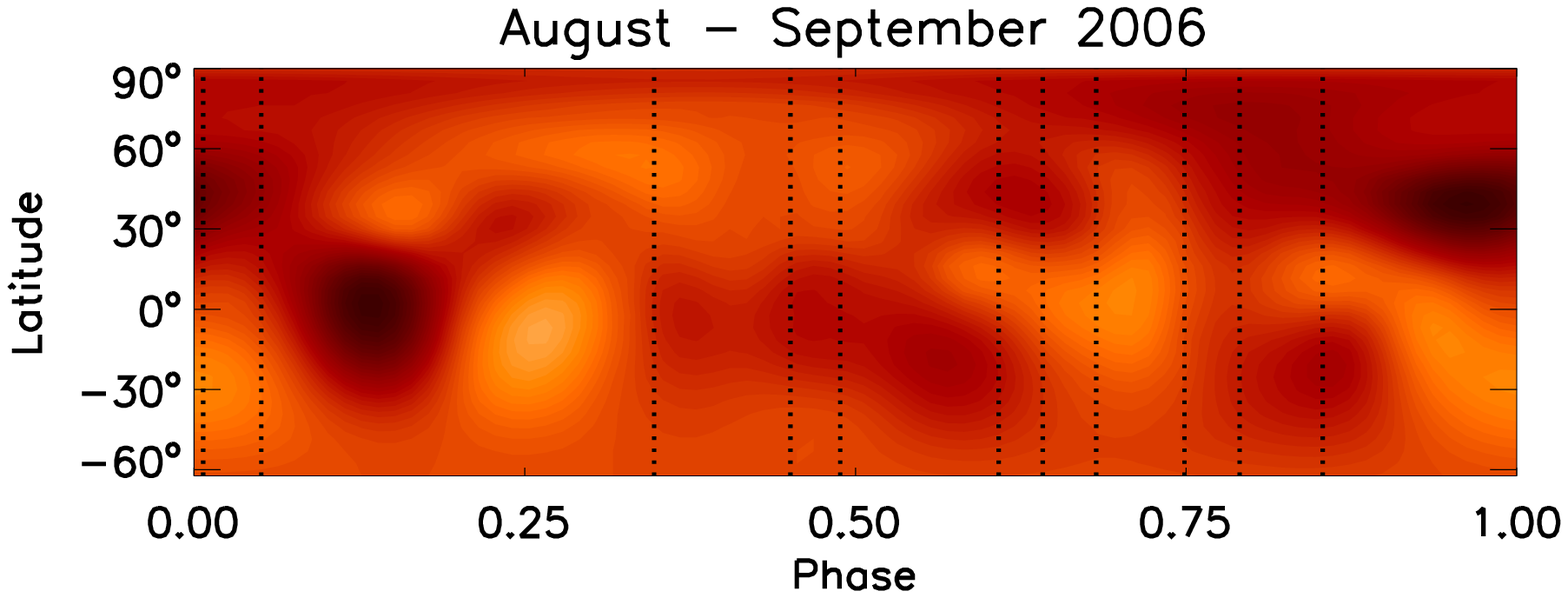}
\includegraphics[width=3.5in]{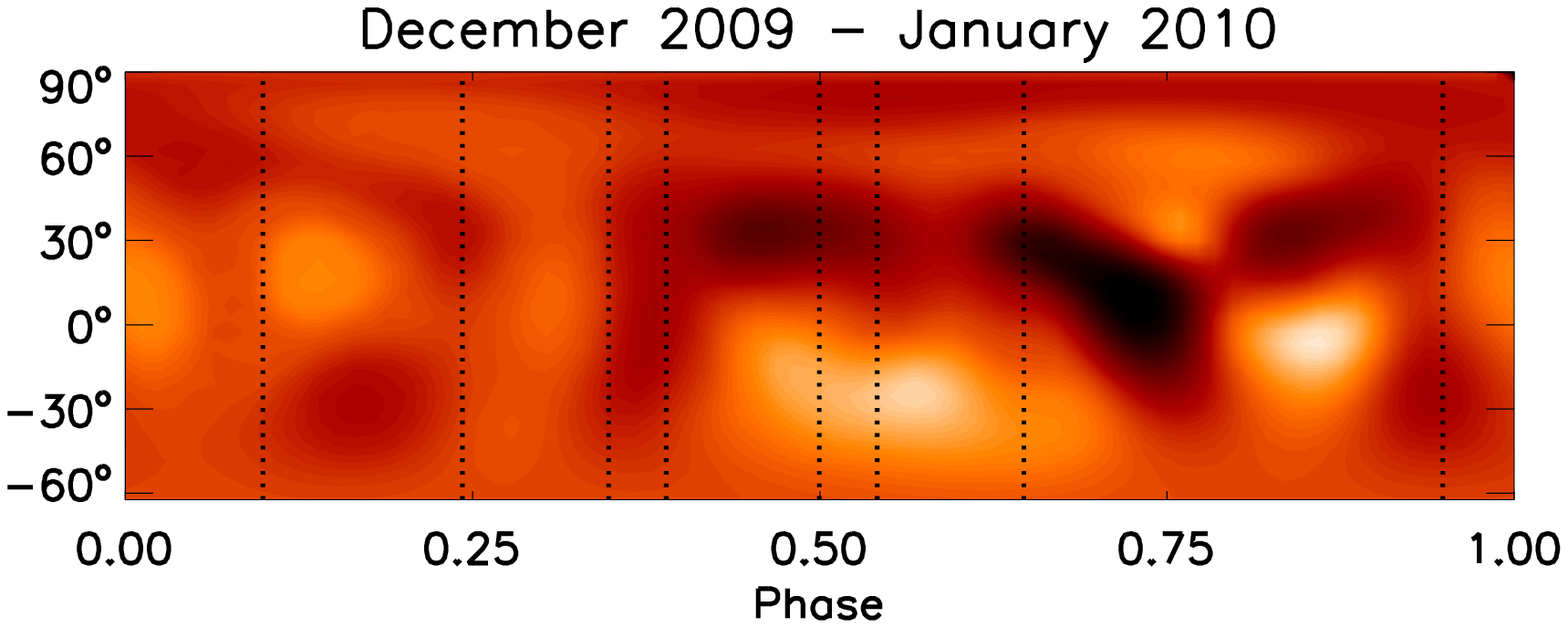}

\vspace{-4.0cm}
\includegraphics[width=2.5in]{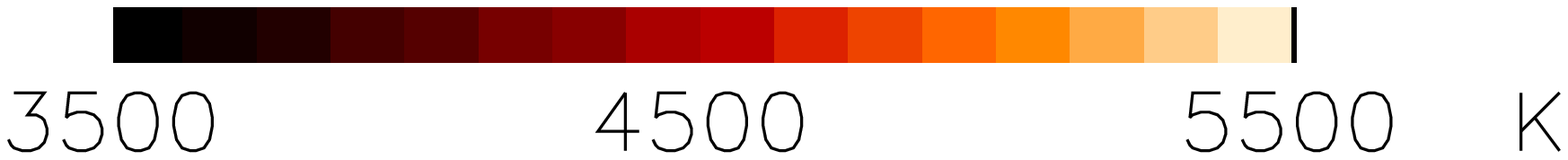}

\caption{Doppler imaging temperature maps from  2002--2010 in cylindrical 
projections. The stellar longitude is given as rotation phase and latitude in
degrees. The phases of the observations are marked with vertical dashed lines.}
\label{dimaps}
\end{center}

\end{figure*}

\begin{figure*}
\begin{center}
\vspace{-2.0cm}
\includegraphics[width=3.5in]{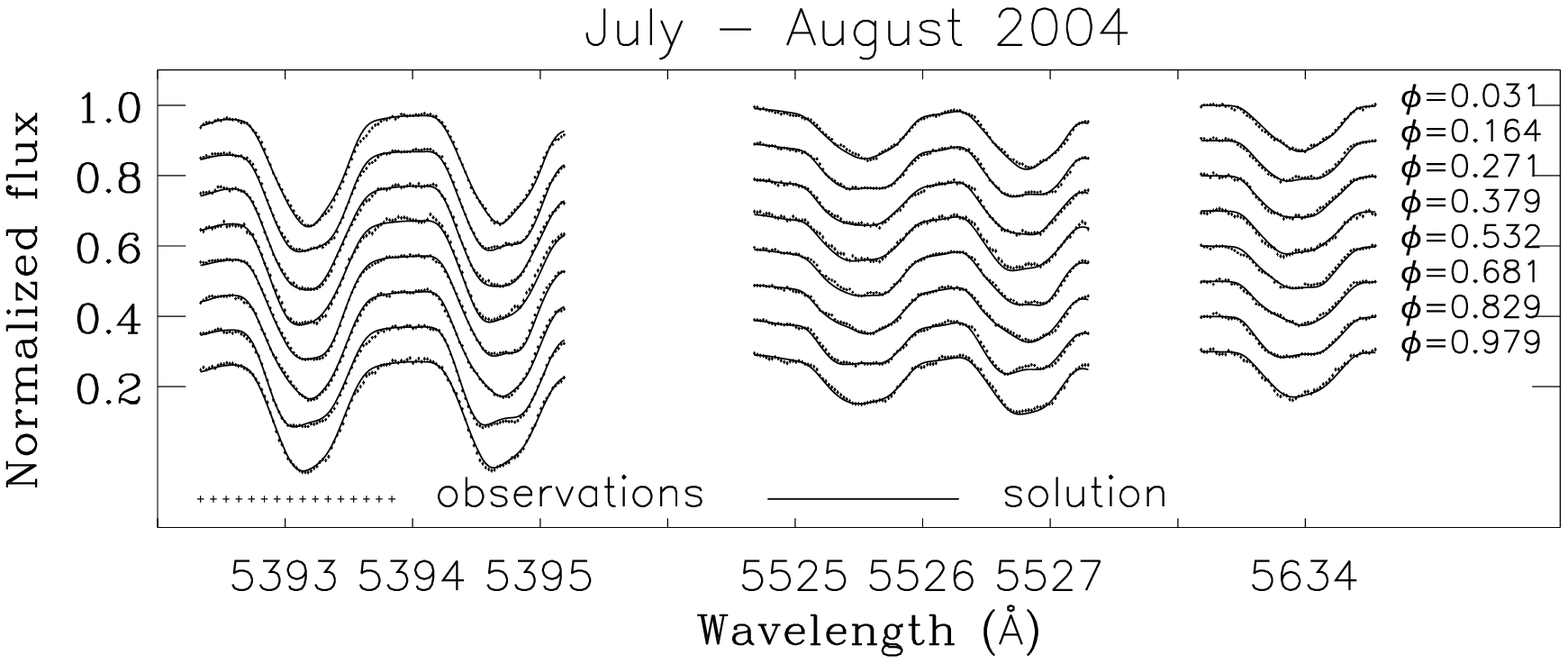}
\includegraphics[width=3.5in]{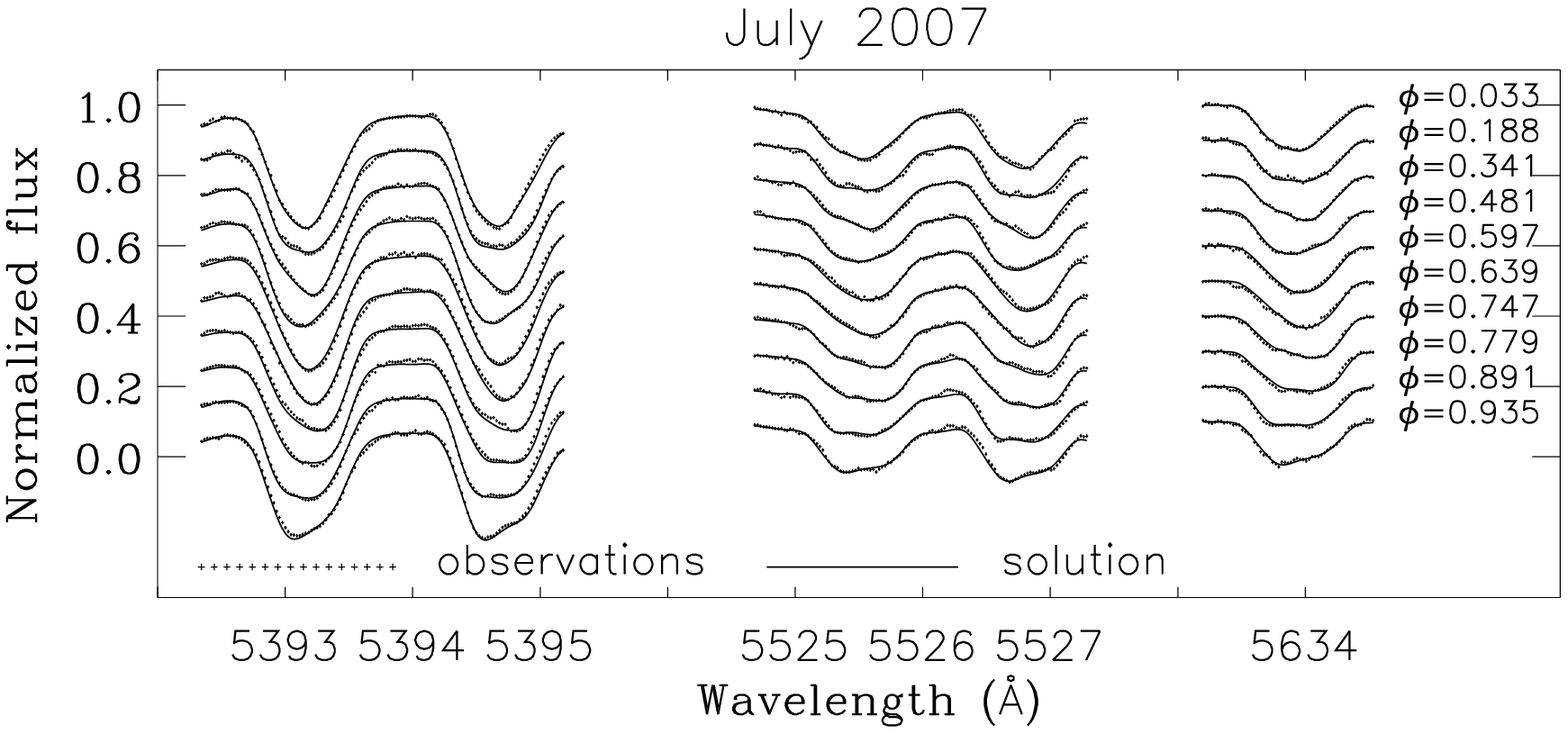}

\vspace{-2.8cm}
\includegraphics[width=3.5in]{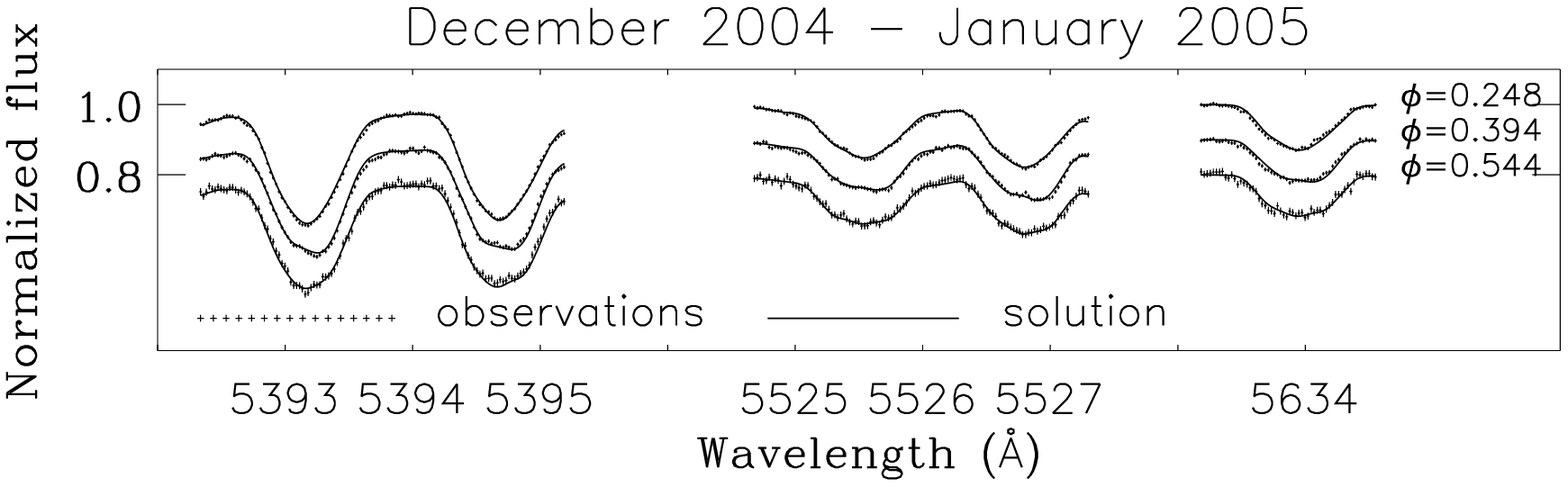}
\includegraphics[width=3.5in]{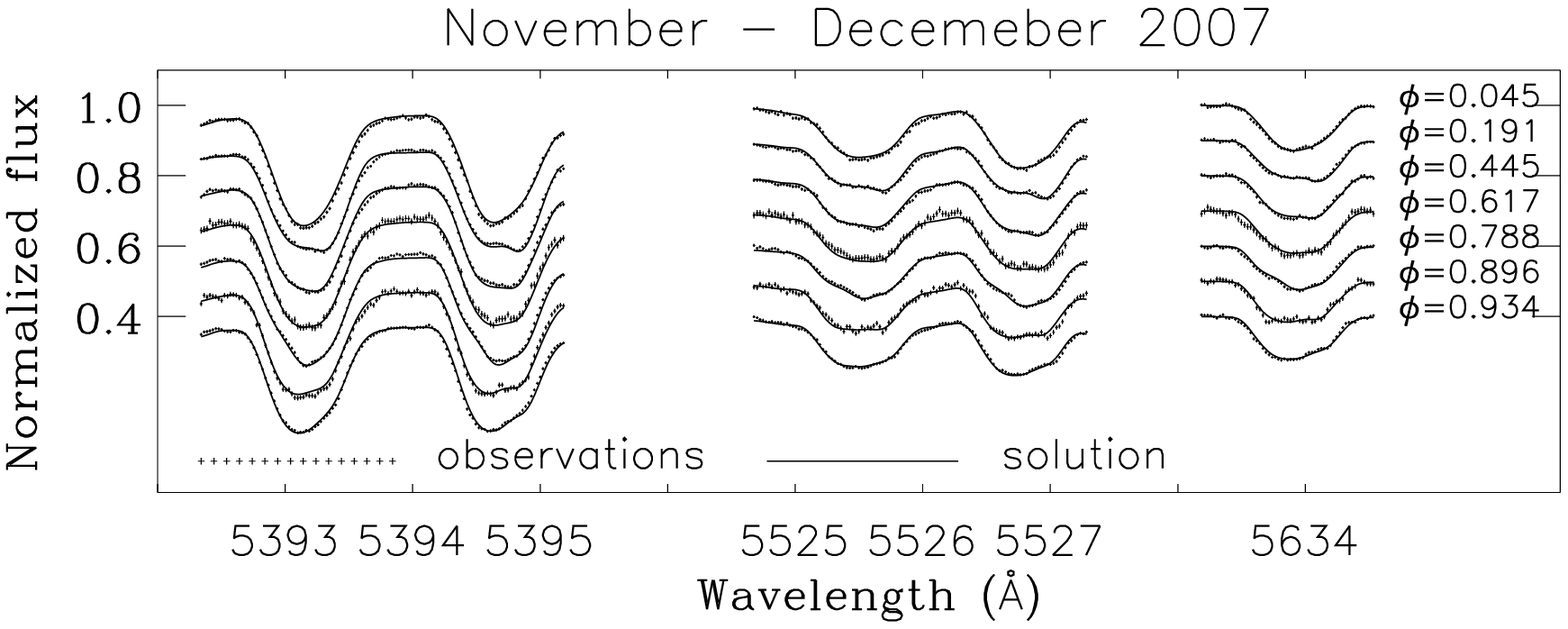}

\vspace{-2.8cm}
\includegraphics[width=3.5in]{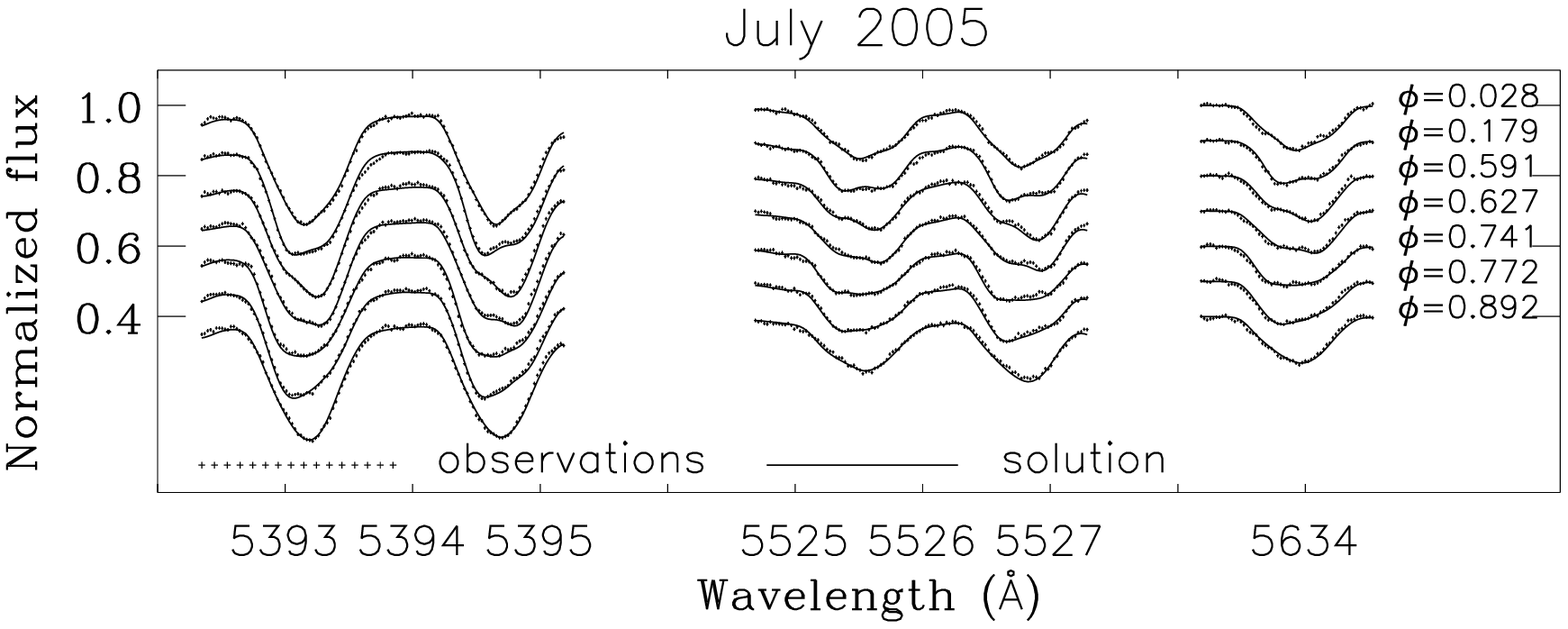}
\includegraphics[width=3.5in]{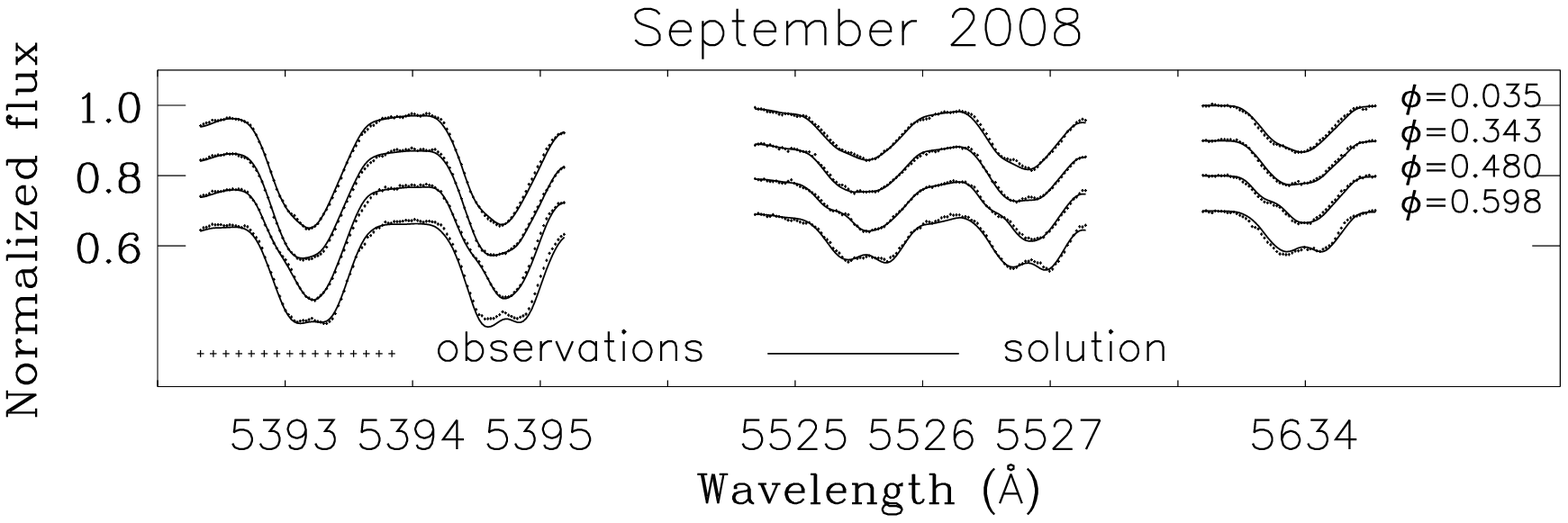}

\vspace{-2.8cm}
\includegraphics[width=3.5in]{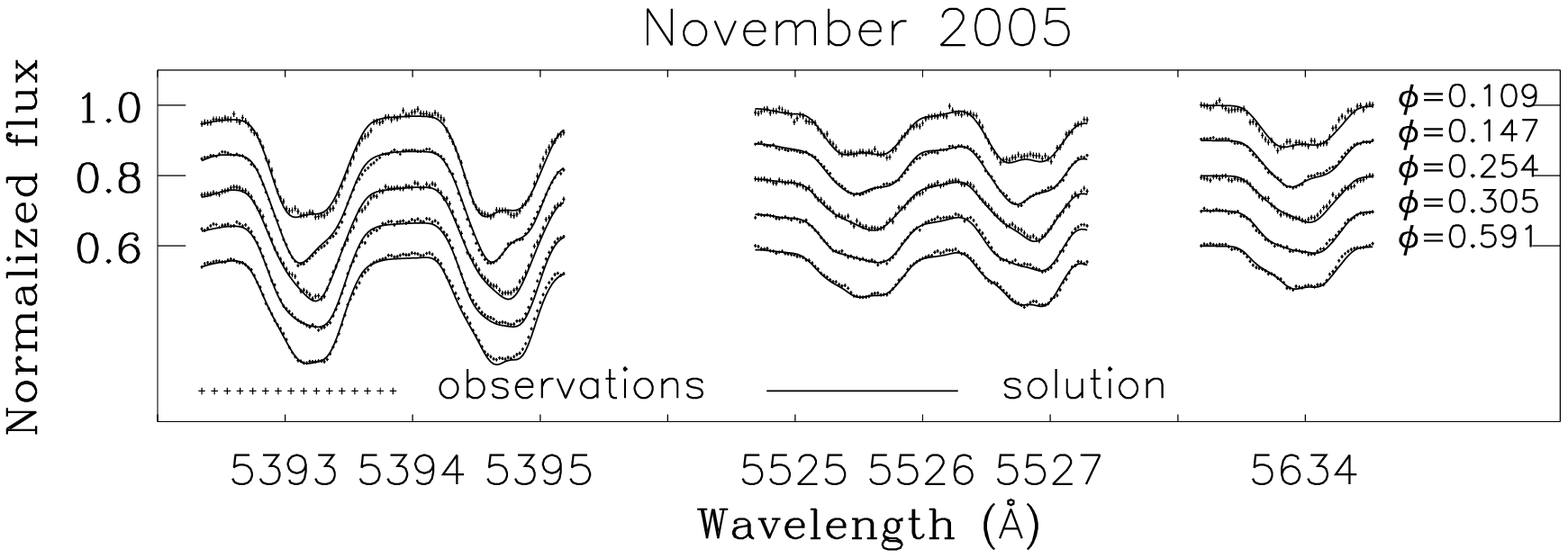}
\includegraphics[width=3.5in]{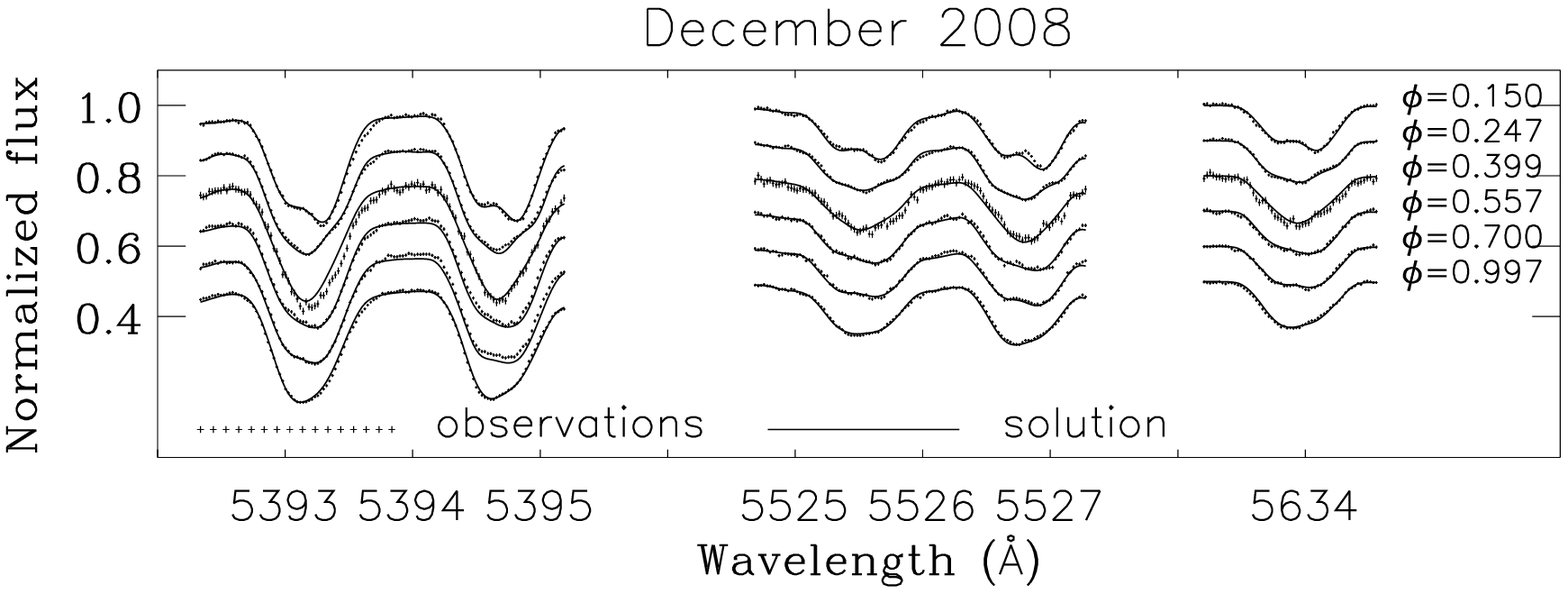}

\vspace{-1.7cm}
\includegraphics[width=3.5in]{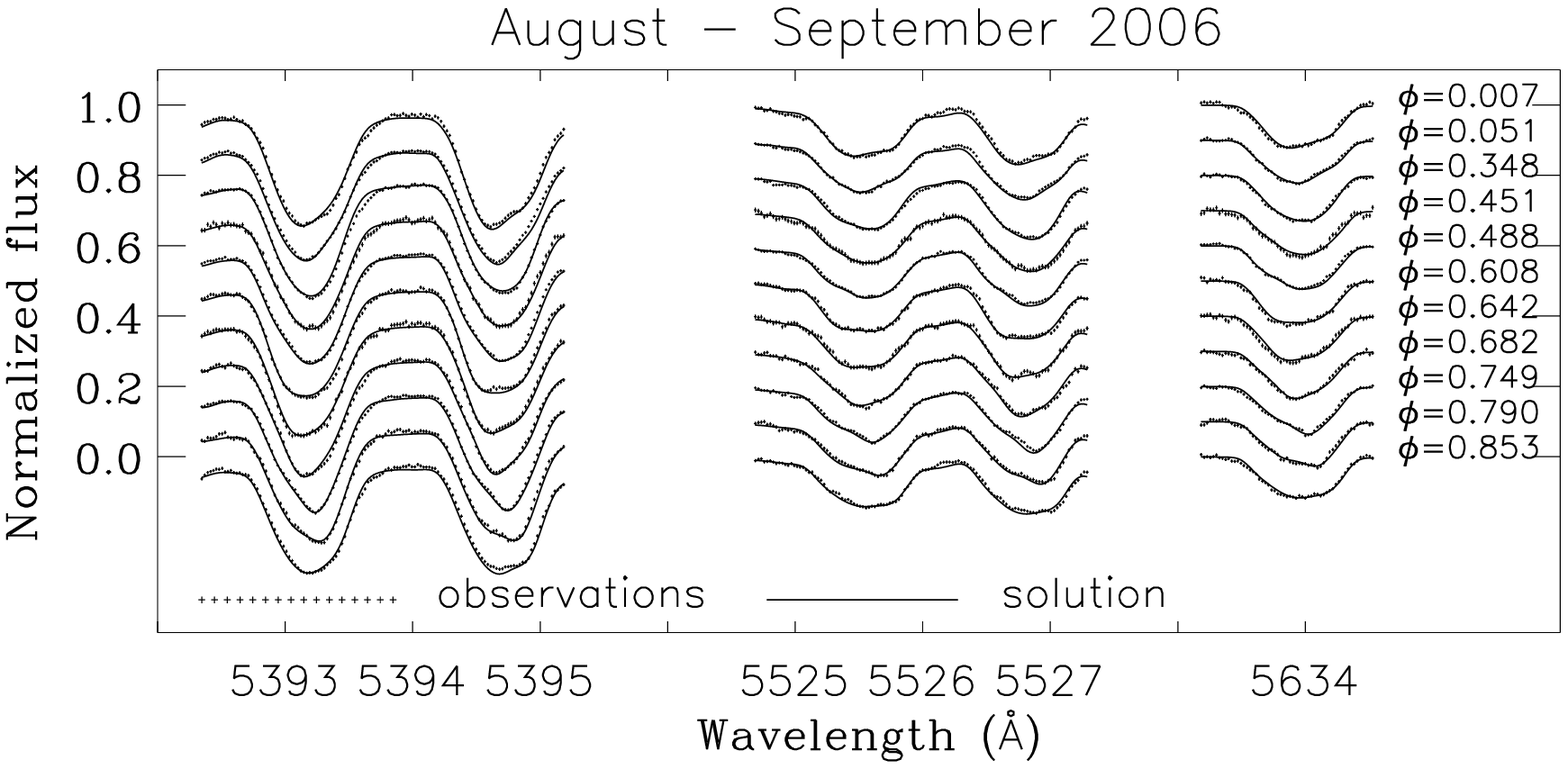}
\includegraphics[width=3.5in]{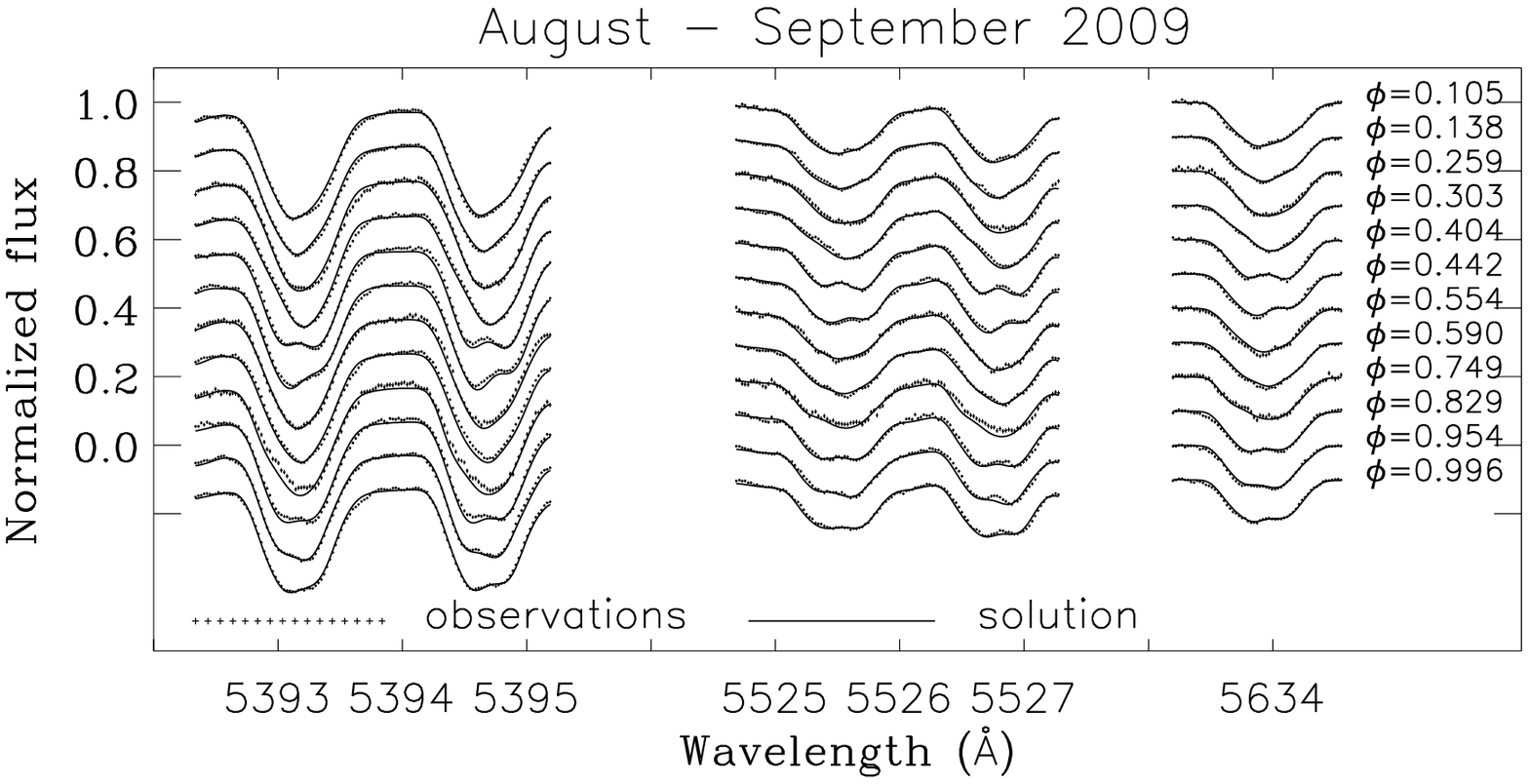}

\vspace{-2.5cm}
\includegraphics[width=3.5in]{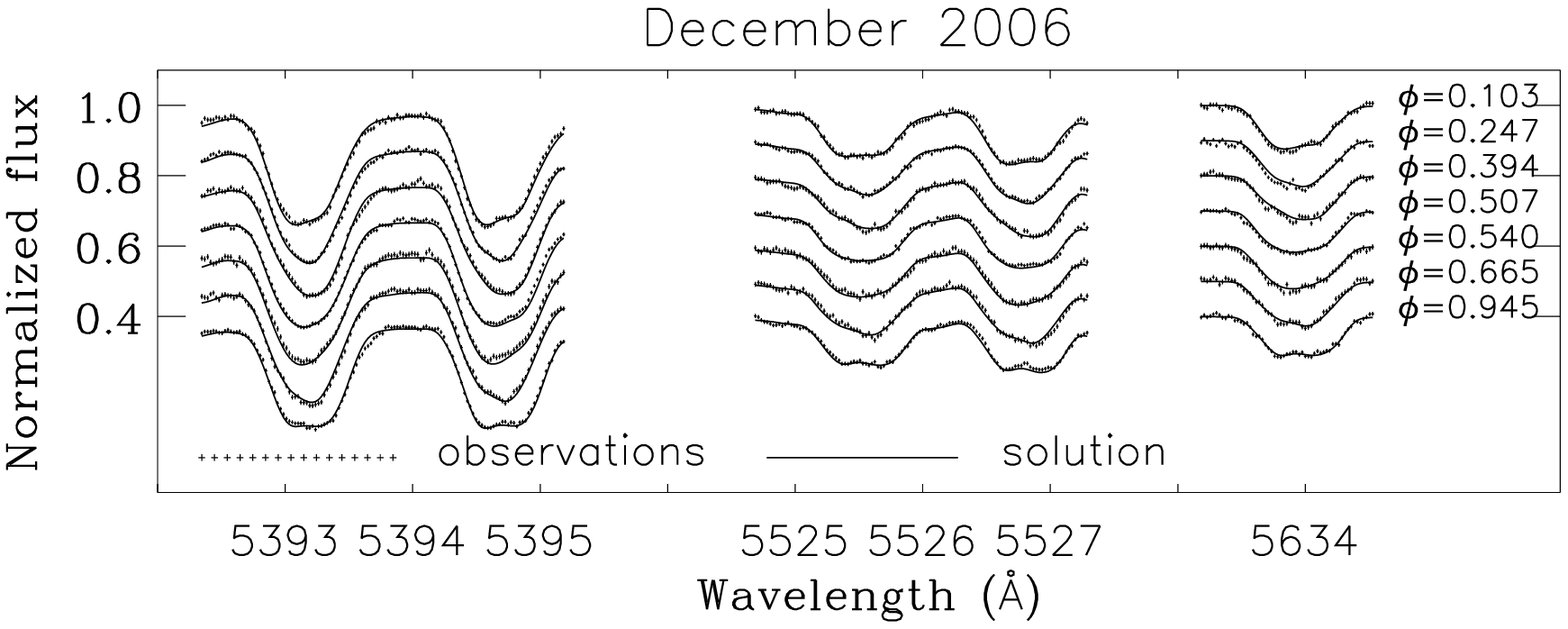}
\includegraphics[width=3.5in]{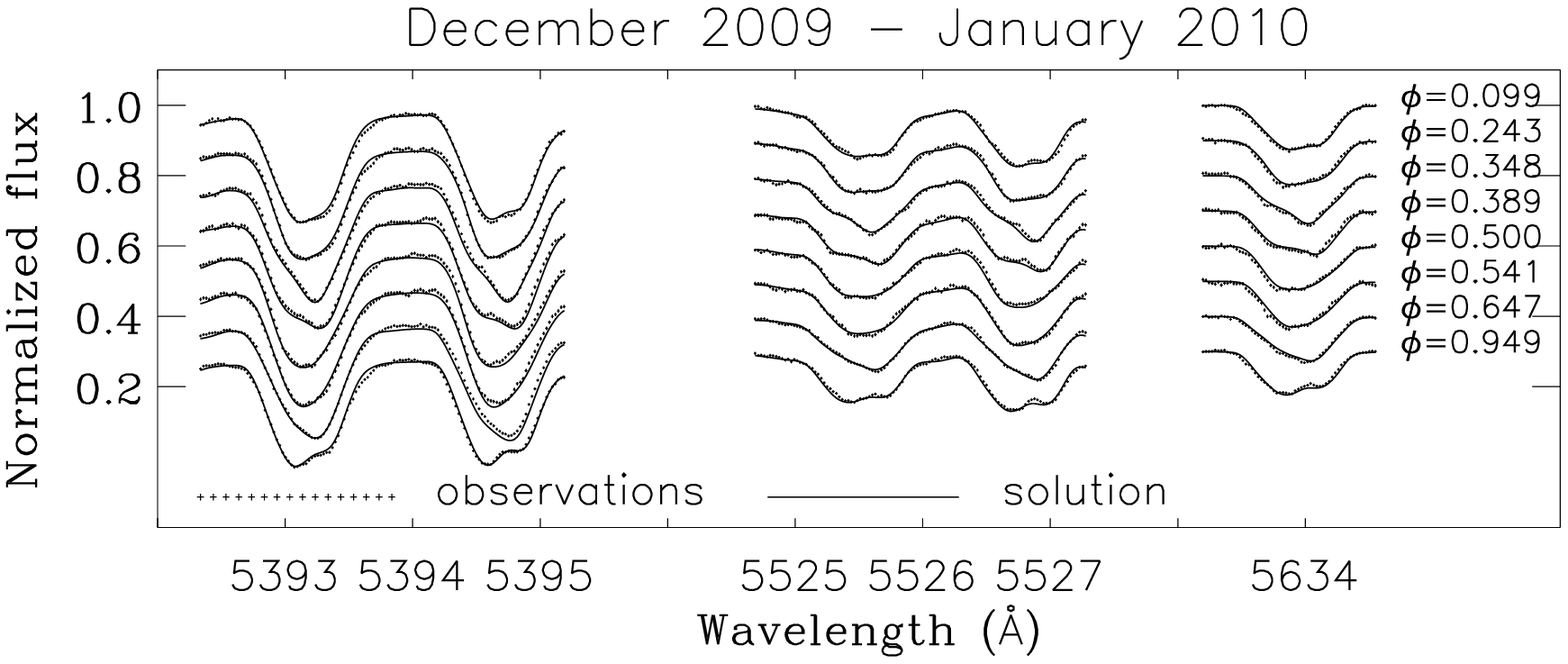}


\caption{The spectral observations and Doppler imaging solutions for
2004--2010.}
\label{dispectra}
\end{center}
\end{figure*}

The inversion was based on Tikhonov regularization. A regularization
parameter of $\Lambda = 1 \cdot 10^{-9}$ and a grid of 80 x 40 surface
elements were used. The aim was to 
achieve a difference between the model and the observations 
of about the same level as the observational noise,
i.e. the inverse of the $S/N$
ratio. This level of convergence was
reached after 40 iterations.  The deviations are listed in
Table~\ref{obssum}.  
In most cases the deviation was significantly larger than the inverse of
the $S/N$-value. This
may be caused by systematic errors, e.g. 
slight shifts in the continuum level or modelling errors. In some 
seasons, there may also be changes in the spot configuration happening on a 
timescale shorter than a week. The
resulting maps are displayed in Fig.~\ref{dimaps} and comparisons
between the modelled and observed spectra are shown in
Fig.~\ref{dispectra}.

To make a more reliable comparison with our earlier Doppler images
of 1994 -- 2002 \citep{lindborg2011}, we recalculated the images from 2002 
using the new stellar parameters and model atmospheres. Since the earlier 
observations did not include the same wavelength regions as the present ones,
we used the same wavelength regions as in the previous study. The resulting maps
are shown in the two uppermost left panels of Fig.~\ref{dimaps}.

\begin{table}
\caption{Stellar parameters.}
\centering
\begin{tabular}{ll}
\hline \hline
Parameter & Value \\
\hline
Gravity  & $\log g =3.5$ (in cgs-units) \\
Inclination & $i = 60$ \degr \\
Rotation velocity & $ v \sin i = 22.6 $ km/s  \\
Rotation period & $P = 6 \fd 724333$\\
Metallicity & $\log [M/H]=-0.25$ \\
Macroturbulence &  $\zeta_{\rm t}=3.5 $ km/s \\
Microturbulence & $\xi_{\rm t}=1.8 $ km/s \\
\hline
\end{tabular}
\label{sparam}
\end{table}

\begin{table}
\caption{Parameters for important lines.$^{\mathrm{c}}$}
\centering
\begin{tabular}{lccc}
\hline \hline
Element ion & $\lambda_\mathrm{centr}$ & $\chi_\mathrm{low}$ 
& $\log(gf)$ \\
&  (\AA) &  (eV) & \\
\hline
\ion{Ni}{I}  & 5392.327 &  4.154 &  $-1.320$ \\
\ion{Fe}{I}  & 5393.167 &  3.241 &  {$\bf -0.865$} \\
\ion{Co}{I}  & 5393.739 &  4.058 &  $-0.326$ \\
\ion{Fe}{I}  & 5394.346 &  4.835 &  $-2.102$ \\
\ion{Mn}{I}  & 5394.677 &  0.000 &  {$\bf -2.803$} \\
\ion{Fe}{I}  & 5394.680 &  4.186 &  {$\bf -1.320$} \\
\ion{Co}{I}  & 5524.985 &  4.113 &  $-0.533$ \\
\ion{Fe}{II} & 5525.125 &  3.267 &  $-4.102$ \\
\ion{Fe}{I}  & 5525.477 &  4.209 &  $-1.994$ \\
\ion{Fe}{I}  & 5525.539 &  4.230 &  {$\bf -1.484$} \\
\ion{Fe}{I}  & 5525.848 &  5.106 &  $-1.574$ \\
\ion{Sc}{II} & 5526.790 &  1.768 &  {$\bf -0.010$} \\
\ion{Fe}{I}  & 5633.946 &  4.991 &  {$\bf -0.400$} \\

\hline
\end{tabular}
\begin{list}{}{}
\item[$^{\mathrm{c}}$] Central wavelength, lower excitation
potential, and adopted $\log gf$-values. The adjusted values are 
marked with a bold font. 
\end{list}
\label{speparam}
\end{table}

\section{Results}
\label{results}

Our 14 Doppler images of II Peg for the years 2002--2010 
are shown in
Fig.~\ref{dimaps}; the color scale of these images is uniform for all
the maps, to facilitate the comparison of the strength of the spot
activity between the different seasons. Our maps show a continuous
evolution of the spot configuration. The temperature of the coolest
spot structure is usually about
3500--4000 K, except for July 2004, when the spot activity 
is quite weak. In December 2004, a clear structure with two spots 
centred at phases 0.5 and 0.6 has emerged, but during this observational
season only three phases were observed.  In the
July 2005, November 2005, and September 2006 images, two or three
larger spot structures can be seen, the longitude of the spots varying 
strongly in
time. During these observing runs, the phase coverage was considerably
better than for December 2004. In December 2006, July 2007, and November 
2007, the spot activity is again weaker. In September 2008 and December  
2009, the Doppler images are dominated by one or two
very cool spots, while the spot activity seems to be low for the
two observing seasons of December 2008 and August 2009. Furthermore,
we note that the deviation between the solution and
observations, relative to the noise of the observations, is slightly
larger for the December 2008, August 2009, and December 2009
images. This may indicate that the spot activity 
has evolved rapidly during these seasons.

\begin{figure*}
\begin{center}
\includegraphics[width=3.5in]{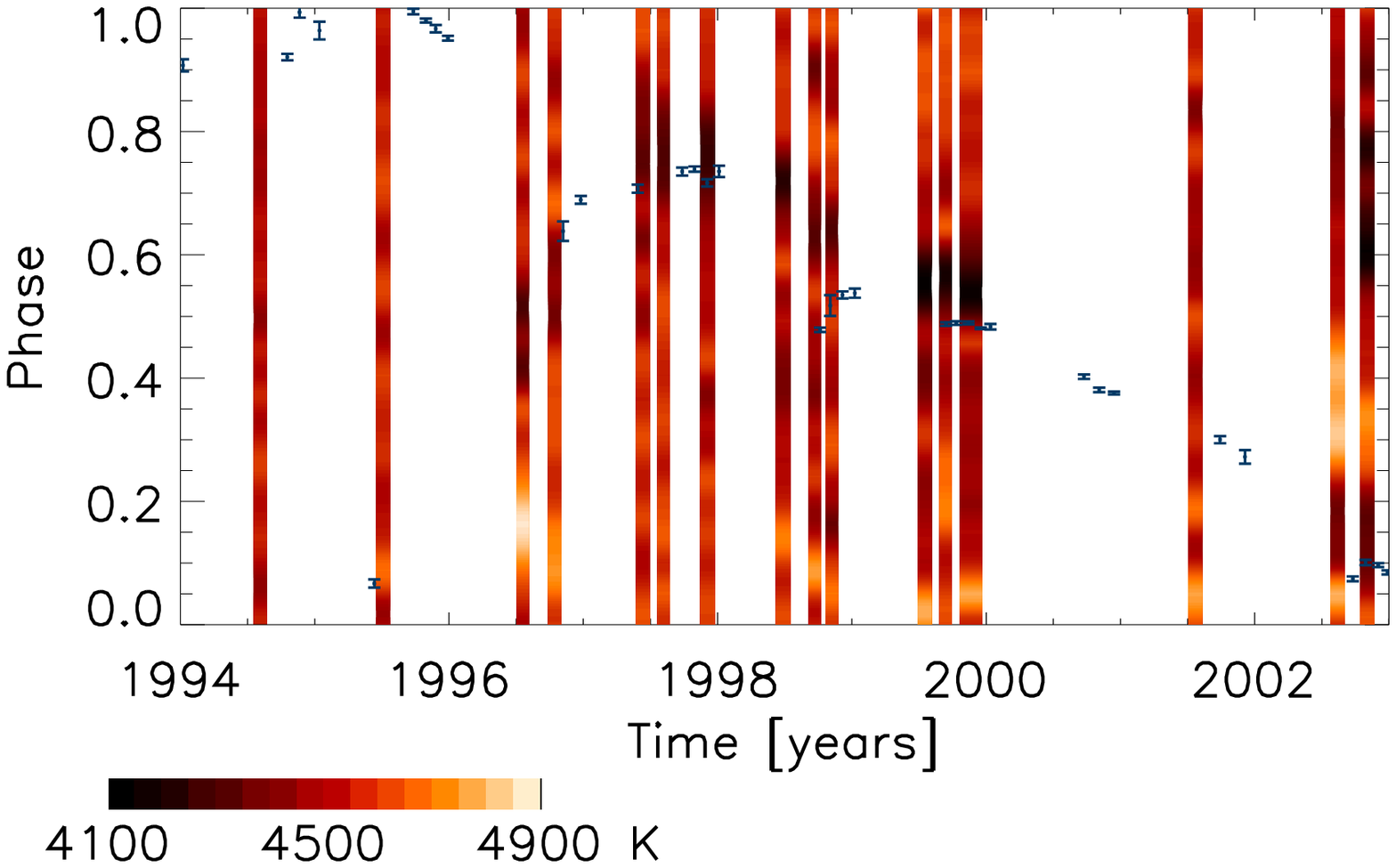}
\includegraphics[width=3.5in]{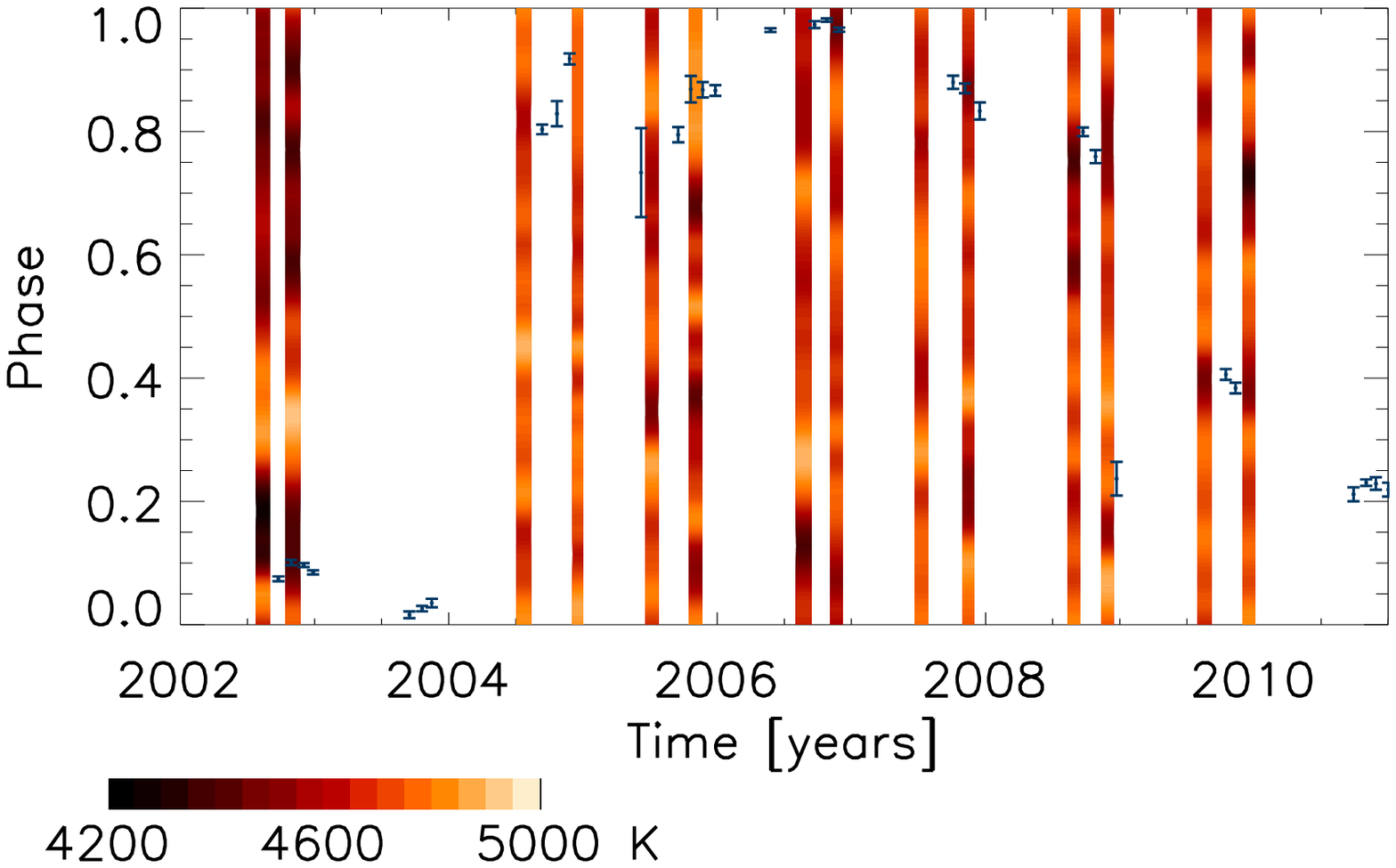}
\caption{The longitudinal spot distribution for the years 1994--2002, 
derived from the old Doppler images \citep{lindborg2011}, and
2002--2010, derived from the images of the present analysis. Each stripe 
represents the temperature  averaged over all
latitudes of the Doppler images. The points with 
errorbars mark the photometric minima.}
\label{phasetime}
\end{center}
\end{figure*}

\begin{figure*}
\begin{center}
\includegraphics[width=3.5in]{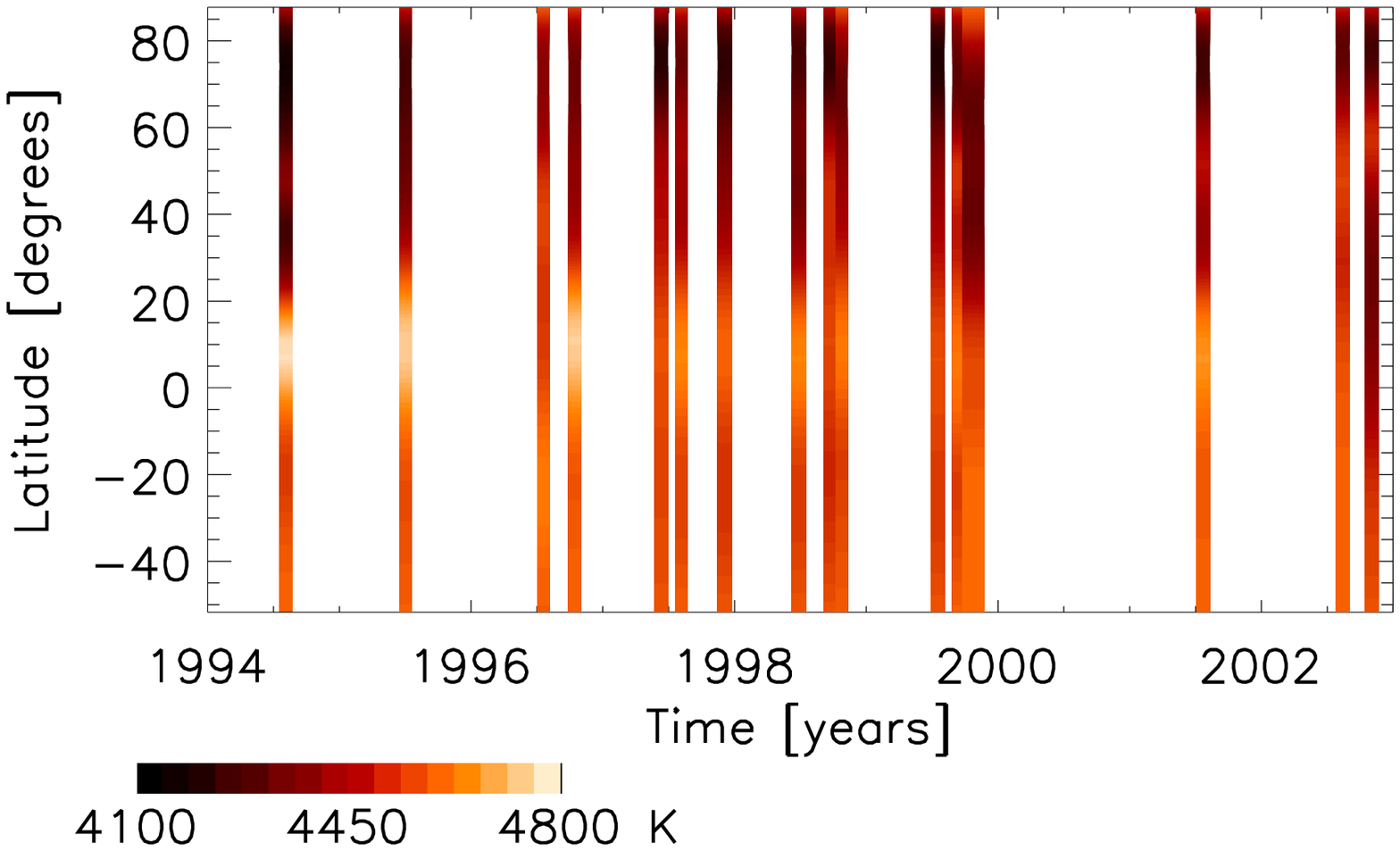}
\includegraphics[width=3.5in]{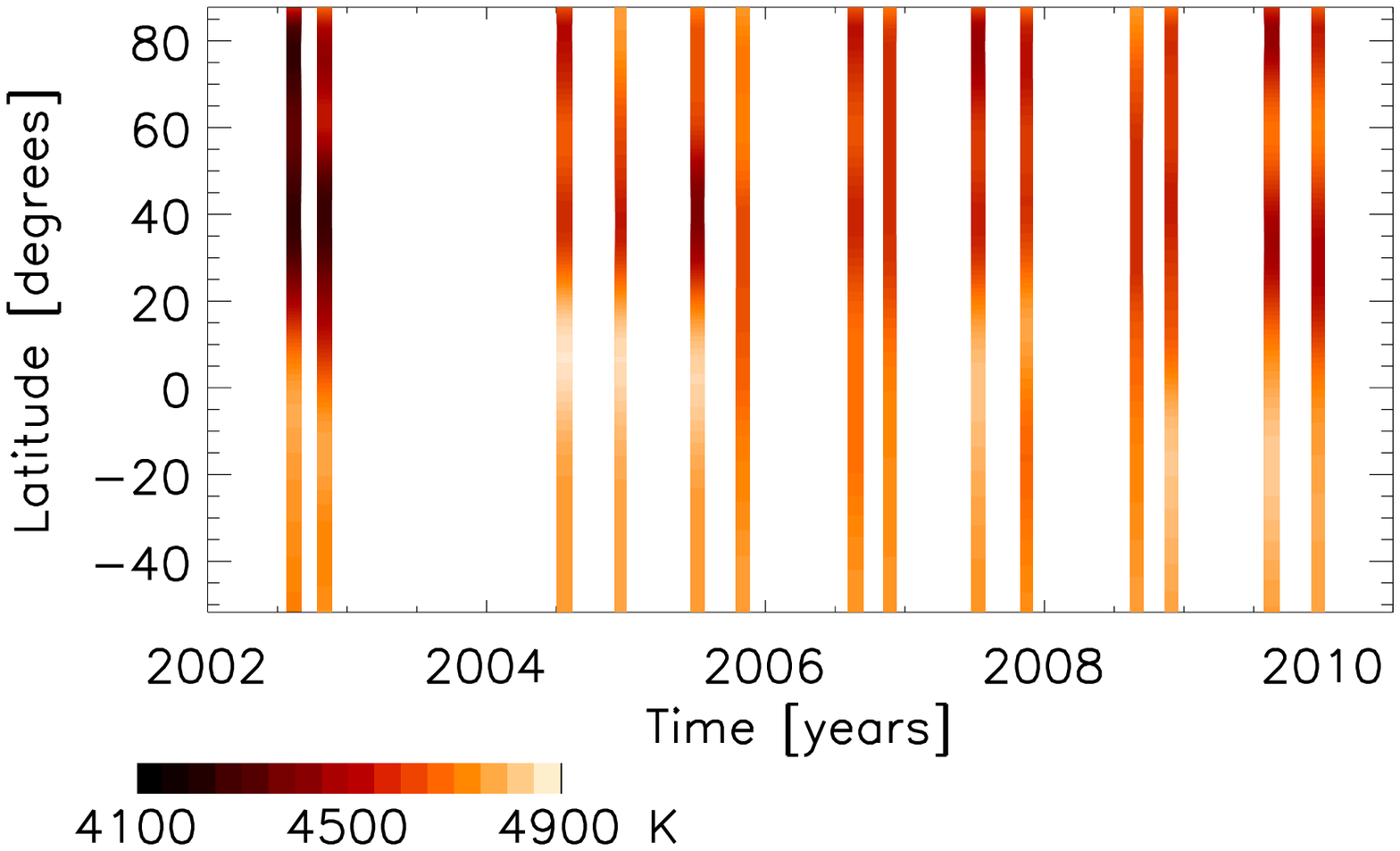}
\caption{The latitudinal spot distribution for the years 1994--2002 and 
2004--2010. Each stripe represents the temperature averaged
over all longitudes of the Doppler images.}
\label{lattime}
\end{center}
\end{figure*}

To study the evolution of the longitudinal spot distribution
over time, we calculated average temperatures over all latitudes for
each image for both our previous Doppler images \citep{lindborg2011}
and the present ones, represented in the left and right panels of
Fig.~\ref{phasetime}. In our previous study of II Peg,
we detected a longitudinal drift of the spot activity in the orbital
rotation frame, which is clearly visible in this plot and also in the maps of
\cite{lindborg2011}. This drift is particularly evident during the
years 1997--1999.  It is hard to detect any such drift in the present
images, especially during 2004--2007, when the spots appear to form at
more or less random phases. During 2008--2010, however, the main spots 
persistently occur on one half, roughly between phases 0.5 and 0.8, of 
the star.

We also performed a similar analysis by averaging the temperature over
longitudes for each latitude. Such a figure could reveal any analogues
to the solar butterfly diagram. However, as can be seen in
Fig.~\ref{lattime}, no clear temporal evolution in the spot latitudes
can be detected. The spots are usually concentrated at latitudes
40--80\degr.  However, this kind of analysis requires
a consistently satisfactory phase coverage, which is not
the case for all the present observations.

We note
that the Doppler images from December 2004, November 2005, and September 2008 
are based on only 3--5 observed phases. 
The longitude of a spot can, in principle, be deduced even from a single 
spectrum, if the spot happens to be near to the centre
line of the visible stellar disk. However, the latitude is determined mainly 
from the amplitude of the radial velocity of the  
``bump'', the measurement of which would require acquisition of
more spectra.

Photometric observations can be used as an additional check of the 
reliability of Doppler images. For this purpose, we used the light curve
amplitude and epochs of photometric minima derived from V-photometry using
the {\it continuous period search} method 
\citep[CPS,][]{lehtinen2011}. 
The analysis was applied to V-photometry obtained with the T3 0.4 APT at 
Fairborn Observatory (Arizona, USA). In the CPS-method,
a sliding window is used to derive continuous estimates
of the light curve parameters. Here we only use the independent measurements
of the amplitude and primary photometric minima, i.e. measurements based
on non-overlapping data. The full
analysis is described by Jetsu et al. (in prep.) and the data are the same
as in the study by \cite{roettenbacher2011}.

In Fig. \ref{phasetime},
we plot both the temperature averaged over latitude of the Doppler images and 
the photometric minima. A basic test of the reliability
of a Doppler image, is that the longitudes of the main spots, 
taking into account their combined effects, coincides with the phase 
of the photometric minimum. In the images of December 2004 and November 2005,
no major spot structures are seen near to the minima derived from 
near-simultaneous photometry. 
The situation with the September 2008 image is somewhat more complicated. In 
the spectra, there is a clear sign, in the form of a bump, of a spot centred
near the rotation phase 0.6 (Fig. \ref{dispectra}). This bump causes the
large cool spot at phase 0.6 in the Doppler image (Fig. \ref{dimaps}). However,
in the nearly simultaneous MOST observations of \object{II Peg} the main 
photometric minimum is near the phase 0.8 \citep{siwak2010}. This is near
the location of the secondary spot in our Doppler image. We can thus conclude 
that  some of the major spot concentrations were misplaced or missed in the 
images from December 2004, November 2005, and September 2008 because of 
insufficient phase coverage.

Even when the phase coverage is good, one should always be cautious not to
over-interpret details of Doppler images. In all our images,
we can identify features that may be artifacts typical of Doppler imaging, 
namely alternating cool and hot regions, arches, and extensions of spots.

The recalculated temperature maps for 2002 (the first two images
in Fig. \ref{dimaps}) are very similar to those
in our previous study \citep{lindborg2011}, the main 
difference being that the average temperature is now $\sim 100-150$ K higher. 
This is a natural 
consequence of the revised stellar parameters in this study. In addition,
there are small shifts in the latitudes of the spot structures, although
the longitudes
coincide very well. The latter is verified by a comparison of the two last
stripes of the left panel with the two first stripes of the right panel in Fig.
\ref{phasetime}. We conclude that the temperature maps of the present study 
can be compared to those
in our previous Doppler images of 1994--2002, but a shift of 
$\sim 100-150$ K in the average temperature should be taken into account.

The mean temperature of the new images for 2002 is still $50 - 100$ K 
lower than that of the images from 2004--2010 (Fig. \ref{activity}). This
could in principle be a bias caused by the change in
the spectroscopic setup. 
The temperature contrast of Doppler images may be affected
by the selection of spectral lines used in the inversion 
\citep[see e.g.][]{jarvinen2010}.
However, a natural explanation
would be that the spot activity has decreased after 2002. This is 
confirmed by the small amplitude of the photometric modulation 
apparent in Fig. 1 of \citet{roettenbacher2011} at around 2004--2005 
(MJD $\sim$ 53000--53500), while it was relatively large at around 2000 
(MJD $\sim$ 51800).

To quantify the level of activity,
we defined all surface elements cooler than
$T_\mathrm{spot} = 4200$ K as spots. The spot coverage of each Doppler image 
was then estimated by calculating the percentage 
of the surface covered by elements with temperatures lower than 
$T_\mathrm{spot}$. There is a clear decrease in the spot coverage, 
which explains the increase in
the mean temperature occurring after 2002 (Fig. \ref{activity}). 
We note that the photometric amplitudes derived by Jetsu et al. (in prep.) also 
support the notion that the spot activity has decreased after 2002.
A comparison with our earlier Doppler images \citep{lindborg2011} also
shows that the spots were more dominant in the images from 1994--2002 than 
during 2004--2010.

\section{Conclusions}

In a recent paper \citep{lindborg2011}, we published 16 temperature maps for the
star \object{II Peg} during 1994--2002, revealing short-term,
irregular, 'flip-flop'-type events and a systematic drift of the active regions
in the orbital reference frame of the binary system. The 12 new
temperature maps for 2004--2010 clearly show that the 
behaviour of the star is quite different from the earlier
epoch:

\begin{figure}
\begin{center}
\includegraphics[width=3.65in]{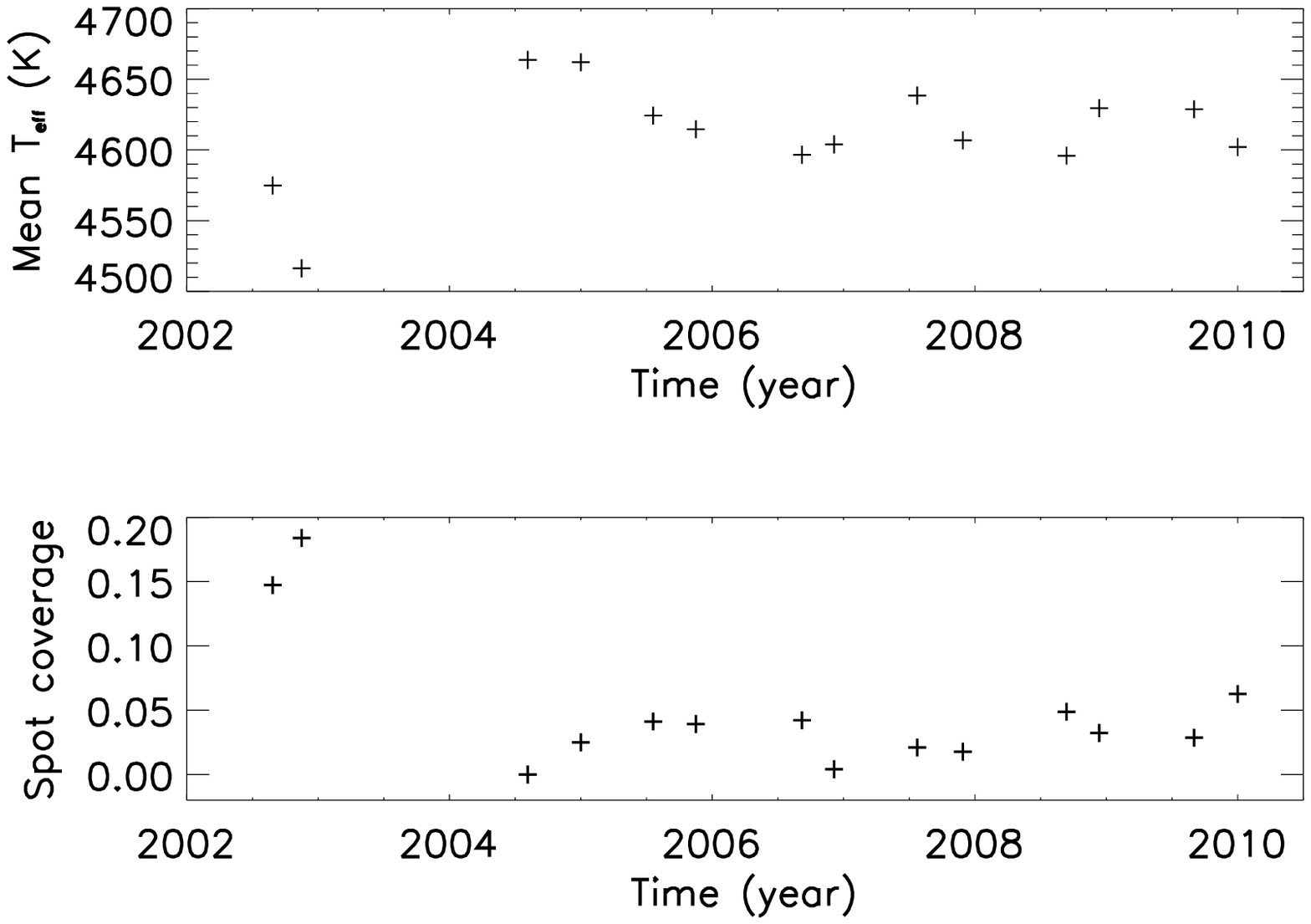}

\vspace{-3.1cm}
\includegraphics[width=3.65in]{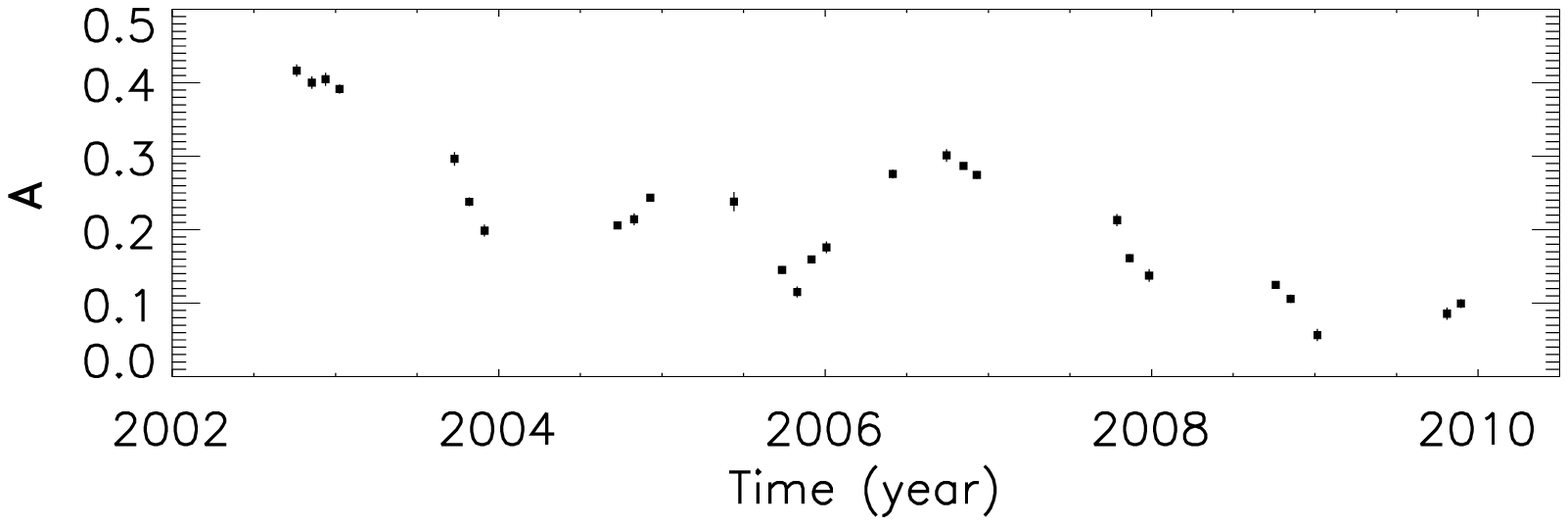}
\caption{The mean $T_\mathrm{eff}$ and spot coverage of the Doppler imaging 
maps, and the photometric amplitude 2002--2010.}
\label{activity}
\end{center}
\end{figure}

\begin{enumerate}

\item 
We find that the spot activity is generally lower than during 1994--2002.
Epochs of low activity, e.g.  July-August 2004, December 2006,
July 2007, December 2008, and August-September 2009, alternate with
states of higher activity. 

\item We are not able to detect any systematic
drift of the active regions from the Doppler images with respect to the 
orbital rotation frame. Instead, consecutive images show 
far less of a resemblance than during the previous
observation period 1994--2002, making it hard to trace any drifts in
the spot structures.

\item There is no evidence of a
'flip-flop' behaviour. Especially during 2004--2007, the spot distribution over
longitudes is more or less random. During 2008--2010, 
large spots persistently occur between the phases 0.5 and 0.8, resembling the
behaviour seen on the object during earlier epochs.
\end{enumerate}

Several studies have found cyclic behaviour in \object{II Peg}. 
\cite{rodono2000} analysed 25 years of photometry and reported cycles of 13.5, 
9.5, and 6.8 years in the spot activity. Furthermore, \cite{berdyugina1999} 
reported a 4.65 year cycle in the 'flip-flop' events.
We cannot confirm any regular flip-flops. The same conclusion
was also drawn in the analysis of photometry by \cite{roettenbacher2011}.

We also see no evidence of a drift in the spot-generating mechanism during
2004--2010 in the Doppler images. However, indications of a drift can be seen
in the photometric minima during 2006--2010. We interpret the drift itself as 
a dynamo wave migrating in the azimuthal direction. 
The spots are thus generated by an underlying structure with a higher angular 
velocity than the surface of the star.

It is clear that the star has
entered a state of weaker activity than during 1994--2002. The spot
evolution seems fast and 
random, which could mean that a dynamo wave cannot be clearly detected. 
This could be related to a minimum in the star's cycle. In this 
respect, the 13.5 year cycle found by \cite{rodono2000} is plausible. A period
of higher activity and a clearly detectable drifting active longitude, 
alternating with a period of lower activity, could then constitute the 
activity cycle. The collected time series
of Doppler images, however, is still too short to make a
decisive conclusion about the existence of such a stellar cycle.

\begin{acknowledgements}
Ilkka Tuominen sadly passed away on March 19, 2011. We wish to express our 
respect for his importance for the research on magnetically active stars.
The work of TH was financed by the research programme ``Active
Suns'' at the University of Helsinki. The work of ML was supported
by the Academy of Finland project 141017, and she has benefited from
the NOT research studentship programme. MJM acknowledges the support from the
Academy of Finland through the project 218159. 
OK is a Royal Swedish Academy of
Sciences Research Fellow supported by grants from the Knut and Alice Wallenberg
Foundation and from the Swedish Research Council. We thank the referee
for valuable suggestions, which helped to improve the paper.
\end{acknowledgements}

\bibliographystyle{aa}
\bibliography{iipeg_fin}

\end{document}